\documentclass[12pt]{iopart}

\usepackage{graphicx}
\usepackage{color}

\usepackage{textcomp}
\usepackage{bm}
\usepackage{amssymb}
\usepackage{cite}

\def\IR{\relax{\rm I\kern-.18 em R}}
\newcommand{\rthree}{{\relax{\rm \kern-.18 em R}}^3}

\begin{document}
\title{Electromagnetics close beyond the critical state: thermodynamic prospect}
\author{A. Bad\'{\i}a\,--\,Maj\'os}
\ead{anabadia@unizar.es}
\address{Departamento de F\'{\i}sica de la Materia
Condensada--I.C.M.A., Universidad de Zaragoza--C.S.I.C., Mar\'{\i}a
de Luna 1, E-50018 Zaragoza, Spain}
\author{C. L\'{o}pez}
\ead{carlos.lopez@uah.es}
\address{Departamento de Matem\'aticas,\\ Universidad de Alcal\'a
de Henares, E-28871 Alcal\'a de Henares, Spain}
\date{\today}
\begin{abstract}
A theory for the electromagnetic response of type-II superconductors close beyond the critical state is presented. Our formulation relies on general physical principles applied to the superconductor as a thermodynamic system. Metastable equilibrium critical states, externally driven steady solutions, and transient relaxation are altogether described in terms of free energy and entropy production. This approach allows a consistent macroscopic statement that incorporates the intricate vortex dynamic effects, revealed in non-idealized experimental configurations. Magnetically anisotropic critical currents and flux stirring resistivities are straightforwardly included in three dimensional scenarios.

Starting from a variational form of  our postulate, a numerical implementation for practical configurations is shown. In particular, several results are provided for the infinite strip geometry: voltage generation in multicomponent experiments, and magnetic relaxation towards the critical state under applied field and transport current. Explicitly, we show that for a given set of external conditions, the well established critical states may be utterly obtained as diffusive final profiles.

\end{abstract}

%
\maketitle

%


\section{Introduction}\label{Sec_1}
Over a half century, the critical state model\cite{bean} (CSM) has been an essential interpretative tool for the investigation of the macroscopic magnetic properties of type-II superconductors. This accomplishment probably relies on a clear physical background together with an ostensible mathematical simplicity, that have enabled a huge number of utilizations both in material characterization as well as in more fundamental studies.

In brief, the CSM postulates that the biased magnetic response of a type-II superconductor is defined in terms of a series of metastable equilibrium configurations, each characterized by a current density distribution ${\bf J}({\bf r})$. In the simplest cases, ${\bf J}({\bf r})$ may be obtained by integration of Amp\`ere's law given by $dH/dx = \pm J_c\, , 0$. $J_c$, the so-called critical current density is the single material parameter of the theory, and characterizes the balance equation between magnetic and intrinsic pinning forces: ${\bf J}\times{\bf B}={\bf F}_p$. The transition between different configurations is assumed to take place instantaneously, which means that external field variations occur slowly enough as compared to the scale established by the material response ($\tau_{\rm ext}\gg \tau_{\rm mat}$). Thus, although it is well known that magnetic flux penetration in the presence of pinning forces happens as an {\em avalanching process}\cite{pla} when the threshold condition is exceeded (i.e.: $J > J_c$), one argues that magnetic diffusion is so fast that the superconductor settles a negligible time in the intermediate resistive states.

In many instances, the CSM approximation is justified, but especially for the case of high $T_c$ superconductors one can meet practical situations for which relaxation transients towards equilibrium configurations have to be considered. A very relevant case is the {\em fault current limiter} that, by construction operates driving the superconductor beyond the resistive transition\cite{surdacki,cha03}. The precise knowledge of the magnetic diffusion processes that occur is obviously necessary for the design of such devices. In particular, the essential evaluation of energy losses may be seriously tampered if one neglects the effect of the sample becoming resistive. On the other hand, and more to the side of basic physics, one has to recall that, experimentally, the value of the critical parameter $J_c$ is often obtained from transport measurements that are based on some threshold value for the voltage detected when the sample goes resistive (typically $1\mu V/cm$). In a recent work about the origin of the dissipation mechanisms that operate for superconductors in the vortex state, it has been argued that, although designed for obtaining the critical parameters, resistive measurements could be ineluctably recording properties of the resistive state\cite{clem}. This circumstance is relevant for HTS samples that display more gradual current voltage transitions as compared to the sharp behavior for conventional superconductors. Finally, it should also be mentioned that from the fundamental point of view, embedding the CSM approximation in a time dependent theory that allows to derive it as a limiting case is by itself a desirable objective.

In this work we put forward a minimal formulation that upgrades the critical state theory so as to enable the inclusion of {\em overcritical behavior}. The basic picture of our statement is as follows. Subject to an external action, the state of the superconductor will drift until new steady conditions are met. If allowed, the steady state will be an equilibrium phase, characterized by a conventional {\em force balance} equation, with critical behavior for the current density. On the other hand, dissipative {\em steady states} and excursions from one critical state to the other will be described under a first order (linear) approximation for the underlying driving forces. Following the spirit of Bean's model, although describing phenomena whose ultimate nature would at least require a mesoscopic scale, we pursue a theory based on macroscopic variables. Thus, our basic assumption is that the superconductor behaves as a thermodynamic system, with the coarse-grained current density as the state variable, and postulate that dynamic properties may be predicted on the basis of three factors: 

\begin{itemize}
\item[(i)] magnetic forces
\item[(i)] material pinning forces
\item[(iii)] irreversible thermal forces. 
\end{itemize}

The latter group will allow to introduce the physical effects related to dissipation induced by overcritical current flow. In order to avoid microscopical modeling, a power series expansion argumentation will be used for introducing such effects. Outstandingly, recalling universal properties as entropy increase and single valuedness of the physical observables, one can fairly determine the kind of models to be used, even in complex scenarios such as flux cutting environments or 3D modeling.

It must be clarified that, being interested in the description of small excursions beyond the critical state, long term effects, such as thermal relaxation to true equilibrium are not considered here. Thus, our theory describes the behavior of superconducting material with a characteristic a time very short as compared to thermal creep relaxation, but non negligible against external source variations, i.e., $\tau _{\rm mat} \lesssim \tau _{\rm ext} \ll \tau _{thermal}$. Therefore, magnetic diffusion times are small but not negligible against external sources variations, while thermal creep (of the order of hours or days) is ignored, determining a well defined metastable equilibrium. Furthermore, isothermal conditions will be assumed and, thus, any energy transfer from the electromagnetic sector would be immediately absorbed by the surrounding thermal bath.

The article is organized as follows. Section \ref{Sec_2} is devoted to deliver the formal details of our theory. First (Sec.\ref{Sec_21}) we recall some mechanical concepts (fields, forces, Drude's model) that allow a rudimentary approach to the problem of dissipation in normal conducting systems. Then, (Sec.\ref{Sec_22}) some basic thermodynamic background is introduced, with the aim of paving the way for the generalization to superconductors. As a central result of our work, in \ref{Sec_222} we introduce the idea of {\em dissipation function} and entropy generation for type-II superconductors driven out of equilibrium. We will show that, concomitant with a complex structure for the metastable equilibrium states, an acceptable theory of resistive losses has to fulfill some consistency requirements, that will be used to put restrictions on the possible material laws to be used (${\bf E}({\bf J})$ in particular). In order to ease the practical implementation, a variational form of our theory is issued. The second part of our paper (Sec.\ref{Sec_3}) illustrates the application of the above concepts to practical situations. Mainly focused on the infinite strip geometry (quasi-1D configuration), we show that our basic equation allows to obtain the electromagnetic quantities either in equilibrium, steady states or during transient processes. It is explicitly shown that, for a given set of external conditions, relaxation eventually leads to the well known corresponding critical states. Excellent comparison with analytical results, when available, is displayed. The final section (Sec.\ref{Sec_4}) summarizes our results and contains a brief discussion of possible applications and extensions of our work in the area of type-II superconductivity.


\section{Resistive losses in hard superconductors}~\label{Sec_2}

\subsection{Coarse grained modeling: fields, forces and ${E}({J})$ law}~\label{Sec_21}

Let us start by recalling some details about the classical description of electrical conduction in normal metals, {\em viz.}, the Drude model. As we will see, this simple scheme may be illuminating for the issue of a minimal model of the resistive behavior of hard superconductors.  

\subsubsection{Normal metals}~\label{Sec_211}

The simplest description of electric current in normal conductors (Drude model) is built from the classical dynamical equation of the charge carriers subject to both an external electric field and a phenomenological drag force ${\bf F}_{\rm drag}$ standing for the interaction with the molecular environment and other charges, i.e.:

\begin{equation}
\label{Eq_drude}
 m_{e}\frac{d{\bf v}_e}{dt} = -e {\bf E} + {\bf F}_{\rm drag}({\bf v}_e) \, .
\end{equation}

Customarily,  ${\bf F}_{\rm drag}$ is taken to first order, i.e.: ${\bf F}_{\rm drag}\approx - (m_{e}{\bf v}_e/\tau_{\rm tr})$ and it is a characteristic of the material through the time constant $\tau_{\rm tr}$.
It is important to recall that all the quantities in the above formulae have to be interpreted as mesoscopic averages both in spatial regions and time intervals. This entails smoothing (by randomizing)  the macroscopically irrelevant fluctuations of the microscopic level. In a thermodynamic language, one speaks about a thermalized electron-lattice system. 
In spite of its simplicity, the model catches the basic behavior observed in an overwhelming amount of experiments. 

It is of particular interest to observe that, unless for very high excitation frequencies (in the range of 1 THz for typical metals), further simplification is possible. First, a quasisteady state approximation may be used, that implies to neglect the time derivative (electron inertia) to the left hand side of Eq.(\ref{Eq_drude}). Then, the non-dispersive form of Ohm's law arises, i.e.: ${\bf J}=(ne^{2}\tau_{\rm tr}/m_{e}{\bf E})\equiv \sigma_{0}{\bf E}$ with $n$ the number of conduction electrons per unit volume. Also, related to the use of good conductors, as it is customary in conventional electric machines, charge accumulation within the sample may be neglected (i.e.: $\partial\rho_{q} /\partial t \approx 0$) and one may use Amp\`ere's law in its {\em magnetoquasistatic} form $\nabla\cdot{\bf J}=0$. In practice, when studying the evolution induced by some process of the external excitation, this means that one starts by calculating the {\em dominant} magnetic fields from the diffusion equation, that for latter convenience will be expressed in terms of the material resistivity $\rho_{0}\equiv 1/\sigma_0$
\begin{equation}
\label{Eq_diff}
 \left(\mu_{0}\frac{\partial}{\partial t}-\rho_{0}\nabla^{2}\right){\bf H}=0 \, 
\end{equation}
Eventually, {\bf E} is obtained from Faraday's law. Notice that {\em stationary} solutions of this equation verify the condition $\nabla\times{\bf J}=0$, that is to say: persistent current loops within the sample are excluded for a normal metal.

\subsubsection{Hard superconductors}~\label{Sec_212}

Microscopically, the nature of resistive losses in flux penetrated type-II superconductors is quite more complex than the scattering of normal electrons, which is behind the above summarized Drude's model. Nevertheless, one can take benefit of the conceptual aspects introduced, that may be straightforwardly applied at the phenomenological level. Thus, it is well known that under the action of a transport current and a perpendicular magnetic field, these materials experience a resistive transition, also characterized by a linear ${\bf E}({\bf J})$ relation, now in terms of the so-called flux-flow resistivity $\rho_{\rm ff}$. This behavior is well understood at the mesoscopic level. Abrikosov vortices are driven by the Lorentz-like force ${\bf J}\times{\bf B}$ and their normal cores contribute to the transport current as a normal channel. Then, in the absence of additional effects, one could formulate the behavior of the superconductor in the same terms described above for a normal metal, just replacing $\rho_0$ by $\rho_{\rm ff}$.

However, the detrimental flux flow behavior has been customarily attacked by a number of pinning strategies. Basically, they rely on the idea of introducing a {\em restoring} force on the vortices, so as to keep them in equilibrium positions while a transport current flows along the system (${\bf J}\times{\bf B}={\bf F}_p$). As pinning forces are bounded ($F_{p}\leq F_{p,max}$), there will be a threshold value for the current density that can flow lossless, the so-called {\em critical current density} $J_c$.
As it was said before, in this framework, the calculations leading to the evaluation of magnetostatic equilibrium properties may be rather simple. Thus, Bean's postulate states that {\em flux penetrated regions will be characterized by the threshold (critical) conditions}, i.e.: $|J| = J_c$. Following this, in 1D problems one just has to integrate Amp\`ere's law in the form $dH/dx=\pm J_{c},0$ supplied by specific boundary conditions for the magnetic field. Nevertheless, the investigation of the transient processes involving the appearance of resistivity may be quite complex, even for the simplest geometries. The reason is that now, effective {\em drag forces} only occur for currents circulating with a density beyond the critical value $J_c$. In other words, one has to deal with a material law of the kind $E=\rho(J)J$ where
\begin{equation}
\label{Eq_rff}
\rho(J)=\left\{
\begin{array}{ll}
0 & \qquad  {\rm if}\; |J|\leq J_{c}
\\
\rho_{\rm ff} &  \qquad {\rm if}\; |J|>J_{c} \, .
\end{array}
\right.
\end{equation}

As a consequence of the non-linearity introduced by the piece-wise constant behavior of $\rho (J)$, one has to deal with noticeable difficulties, as for instance, the form taken by the diffusion equation (compare to Eq.(\ref{Eq_diff}))

\begin{equation}
\label{Eq_diff_hard}
 \left(\mu_{0}\frac{\partial}{\partial t}-\rho({\bf J})\nabla^{2}\right){\bf H}=(\nabla\times{\bf H})\times\nabla\rho({\bf J}) \, 
\end{equation}

Difficulties for solving this equation by analytical methods have been reported in Refs.(\cite{surdacki,cha03}) that basically suggest to apply {\em ad hoc} approximations when justified by the set of experimental data under consideration.
On the other hand, we stress that, in fact, the limiting Bean's approximation means to avoid such equation under the following thoughts. If  the actual material law in Eq.(\ref{Eq_rff}) is such that one can speak about a sharp transition, i.e.: $\rho_{\rm ff}$ takes elevated values as compared to  the experimental parameter $E/J$, one can approximate $\rho_{\rm ff}\to\infty$. Then, the magnetic diffusion time ($\tau_{\rho}\propto 1/\rho$) may be neglected and externally induced evolutions may be accurately described as a series of equilibrium {\em critical} states.

%
\begin{figure*}
\begin{center}
{\includegraphics[width=0.65\textwidth]{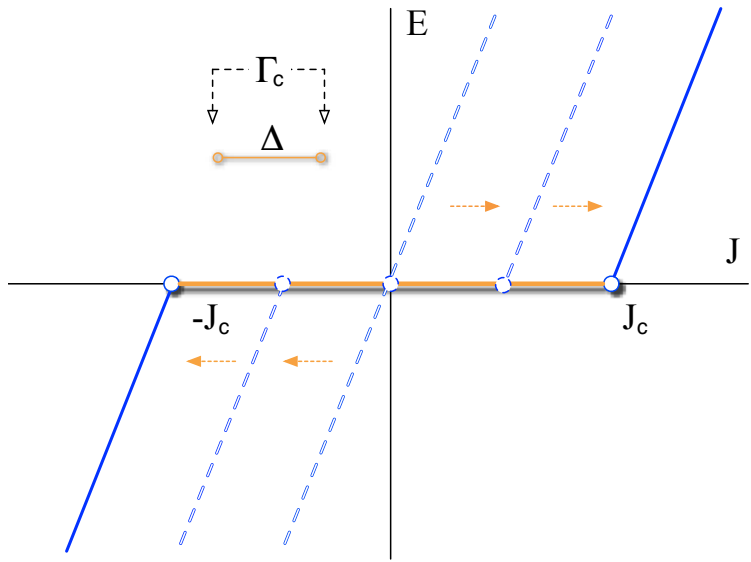}}
\caption{\label{Fig_1}(Color online) Piecewise linear $E(J)$ law for a type-II superconductor. The material parameter $J_c$ defines the boundary $\Gamma_c \equiv\{-J_{c},J_{c}\}$ of the lossless behavior (region $\Delta\equiv [-J_{c},J_{c}]$ on the horizontal axis and detailed as an inset). Ohm's law corresponds to $J_{c}=0\Rightarrow\Delta=\emptyset$. The arrows indicate the effect of introducing pinning forces, and thus progressively increasing $J_c$.}
\end{center}
\end{figure*}
%

Fig.\ref{Fig_1} displays the main features of the above mentioned modeling with a finite value of $\rho_{\rm ff}$ for the case of 1D systems. Let us analyze the physical significance of such a material law.

\begin{itemize}

\item[(i)] Note first that superconducting lossless transport is allowed by the condition $E=0$ within the region $\Delta \equiv [-J_{c},J_{c}]$. The superconducting range $\Delta$ increases with the value of $J_c$, starting from $0$ in very low pinning materials, that behave like ``normal conductors'' (see arrows in Fig.\ref{Fig_1}).

\item[(ii)] Persistent shielding is also possible, subsequent to electromagnetic induction because $E$ can go to zero and $J$ take a local structure that allows $\nabla\times{\bf J}\neq 0$ when diffusion stops. In particular, this includes the critical state solutions with $J$ taking the values $\pm J_c$ or $0$ within the sample, and forming screening loops.

\item[(iii)] One can visualize Bean's model as a limiting case of this material law when there is an arbitrarily high slope of $E(J)$ for $J>J_{c}$.

\item[(iv)] The conceptual implications of Fig.\ref{Fig_1} may be straightforwardly generalized to higher dimensions. In particular, $\Delta$ may be a region of the ${\bf J}$-space and $\Gamma_c$ its boundary. Thus, for the case of high--T$_c$ superconductors it has been recently shown \cite{clem} that a meaningful selection would be to take $\Delta$ as an elliptic region, within the plane defined by the components of ${\bf J}$ parallel and perpendicular to the local magnetic field. Other possible regions, related to specific materials, will be commented on.

\item[(v)] The material law for a type-II superconductor depends not only on the intrinsic parameter $\rho_{\rm ff}$, but is strongly affected by extrinsic quantities. In particular, the ambient magnetic field will play a prominent role for the definition of the region $\Delta$. Thus, already in the simple 1D cases, the dependence $J_{c}(H)$ is usually an important concern. On the other hand, as mentioned above, the relative orientation of the vectors ${\bf J}$ and ${\bf H}$ is crucial for the material law in higher dimensions.

\end{itemize}

\noindent Performing calculations within the above described material law and its {\em natural} extensions to 3D is by no means a simple task. Below, we suggest a first approximation to the problem that relies on thermodynamic concepts applied either close to equilibrium or close to steady states. This will provide a useful framework for developing numerical tools that allow to analyze the resistive transition in applied superconductivity.

\subsection{Thermodynamic background}~\label{Sec_22}
Non-equilibrium thermodynamics is still an area under development. One of the main fields of interest, i.e.: the study of steady states will guide us through the analysis of electrical conduction. In fact, being concentrated on quasisteady systems, the extension to irreversible transient processes (magnetic diffusion) will be straightforward. As an example of what can be done, we recall\cite{landau} that based on the law of increasing entropy and also on Onsager reciprocal relations, one can show that in normal conductors, the conductivity tensor must be positive definite and symmetric. This generalizes 1D Ohm's law. Obviously, such properties could never be deduced from the pure electromechanical principles used in the previous paragraph. Thus, having the aim of a theory that generalizes Sec.\ref{Sec_212} for type-II superconductors, in the forthcoming, we present a minimal conceptual  scenario that allows to host the main physical properties of these materials. As before, we will start introducing the main ideas by taking benefit of our knowledge of the simpler normal conductors.

\subsubsection{Entropy and dissipation function in normal metals}~\label{Sec_221}

Let us start by recalling some definitions. Being interested in local properties of the conductor, we consider the {\em mesoscopic} entropy function ${\cal S}$ as an average that has smoothed the statistical microscopic fluctuations. In fact, the spatial dependence comes through the current density, that will be our {\em state variable}: ${\cal S}[{\bf J}({\bf r})]$. Here, ${\bf r}$ denotes the position of a region of mesoscopic size, and nonlocal correlations are neglected.
In the absence of an external action the system settles at the entropy maximum, i.e.: the stable equilibrium point ${\bf J} = 0$. Physically, an overwhelming amount of all possible microstates corresponds to macrostates with ${\bf J} \simeq 0$, and then ${\cal S}({\bf J}) \leq {\cal S}({\bf J} = 0)= {\cal S}_{\rm eq}$. Statistical fluctuations around such point become negligible with increasing size of the subsystem. 

Out of equilibrium, but not far from it, the same statistical mechanism that suppresses fluctuations will operate. Thus, after a displacement caused by an external agent (electric field in our case), a thermodynamic ``restoring force'' ${\bf F}_{\rm drag}$ in Sec.\ref{Sec_211}, drives the system along increasing entropy until ${\cal S}_{\rm eq}$ is reached again.  ${\bf F}_{\rm drag}$ is responsible for the energy losses that may be expressed in terms of the so-called dissipation function\cite{goldstein} ${\cal F}\equiv(1/2){\bf F}_{\rm drag}\cdot{\bf v}_{\rm e}$, that is a measure of the amount of heat generated per unit time. Macroscopically, the amount of heat generated is given by\cite{explanation}

\begin{equation}
dQ= Td{\cal S}= 2{\cal F} dt \, .
\end{equation}

As stated before, isothermal conditions will be assumed, and we will use the relation $\dot{\cal S}=2{\cal F}/T$. Now, thermodynamics enters through the so-called ``principle of minimum entropy production" \cite{prigogine,klein} that may be expressed as

\begin{itemize}

\item[] {\em The steady state of a system is that state in which the rate of entropy production has the minimum value consistent with the external constraints.}

\end{itemize}

Notice that, based on this, the equilibrium state takes its natural place when there are no constraints on the system, because as a consequence of the second law, it reaches the maximum entropy, and thus the absolute minimum of entropy production, i.e.: zero.

In order to see how the above principle operates, let us assume that some external action (electric field) drives the system out of equilibrium. If the constraint is removed, a transient towards equilibrium (${\cal S}_{\rm eq}$) will occur. Around this point, the function ${\cal F} $ will allow a quadratic expansion of the kind
\begin{equation}
\label{Eq_Omega}
{\cal F}\simeq \frac{1}{2}\sum_{ij} J_{i}\,\Omega^{ij} J_{j} \, .
\end{equation}
with $\Omega^{ij}$  the components of a symmetric positive definite tensor. It is apparent that, in the absence of constraints, minimizing ${\cal F}$ leads to the condition ${\bf J} = 0$, and that the thermodynamic field driving the system towards equilibrium may be obtained as ${\bf E}_{\rm ther} =-\nabla _{\bf J}{\cal F}=-\Omega{\bf J}$. Its lines of force (maximal slope of ${\cal F}$) in the ${\bf J} $-space correspond to the most probable macroscopic evolution (faster increase of entropy). In steady situations, this field {\em balances} the applied electric field that verifies Ohm's law (-${\bf E}_{\rm ther} ={\bf E}= \Omega{\bf J}$). 

On the other hand, steady states out of equilibrium will be characterized by a constrained minimization problem, in which ${\cal F}$ has to be {\em augmented} by some Lagrange multiplier.  For instance, as indicated by Landau\cite{landau}, the steady current distribution within a normal conductor may be obtained by minimizing the volume integral of ${\cal F}+\lambda \nabla{\bf J}$ as corresponds to the stationarity condition $\nabla{\bf J}=0$.

As a main conclusion from the above paragraphs, we resolve that the behavior of a normal conductor close to equilibrium may be formulated in terms of a quadratic tensor $\Omega^{ij}$ (resistivity) that is positive definite and symmetric as dictated by thermodynamics. This obviously generalizes the one dimensional results in Sec.\ref{Sec_211}. An additional observation is that just recalling the definition of the gradient function one can show that the electric field lines will be perpendicular to the constant level sets of the dissipation function ${\cal F}$. Finally, variational principles allowing to afford complicated calculations as those related to the diffusion towards the {\em equilibrium point} (${\bf J}=0$) or characterizing steady sates may be stated in this realm.

Below, we show that the ideas in this section may be exported to the case of type-II superconductors. However, already related to the generalization of the one dimensional $E(J)$ law in Fig.\ref{Fig_1}, some technical difficulties arise, and further analysis is required.

\subsubsection{Entropy and dissipation function in hard superconductors}~\label{Sec_222}

In the light of Sec.\ref{Sec_212} the concept of thermodynamic equilibrium has to be reconsidered for pinned superconductors. Thus, even for the simplest one dimensional configurations, one finds a vanishing drag force not just for a point (${\bf J}=0$) but within a whole segment (region $\Delta$) that becomes a surface, or even a volume in the $J-$space for more complex scenarios. Then, the entropy of the system will be maximal within a full set of values ${\bf J}_{c}\in\Delta$  and the determination of the state from thermodynamic arguments seems flawed by ambiguity. In fact, a complex multiplicity of possible states would seem admissible. Against this, one could argue that, in fact, either static or stationary configurations always occur subsequent to some diffusion process accompanied by drag forces (electric fields) and that ``integration" along some path within the ${\bf J}$-space that connects the overcritical region and the eventual critical point ${\bf J}_{c}$ can be performed. However, the following mathematical quiz arises: how does one expand a function ${\cal F}$ towards the resistive behavior when a ``starting region", instead of a starting point, is given? 

\begin{figure*}[h]
\begin{center}
{\includegraphics[width=0.98\textwidth]{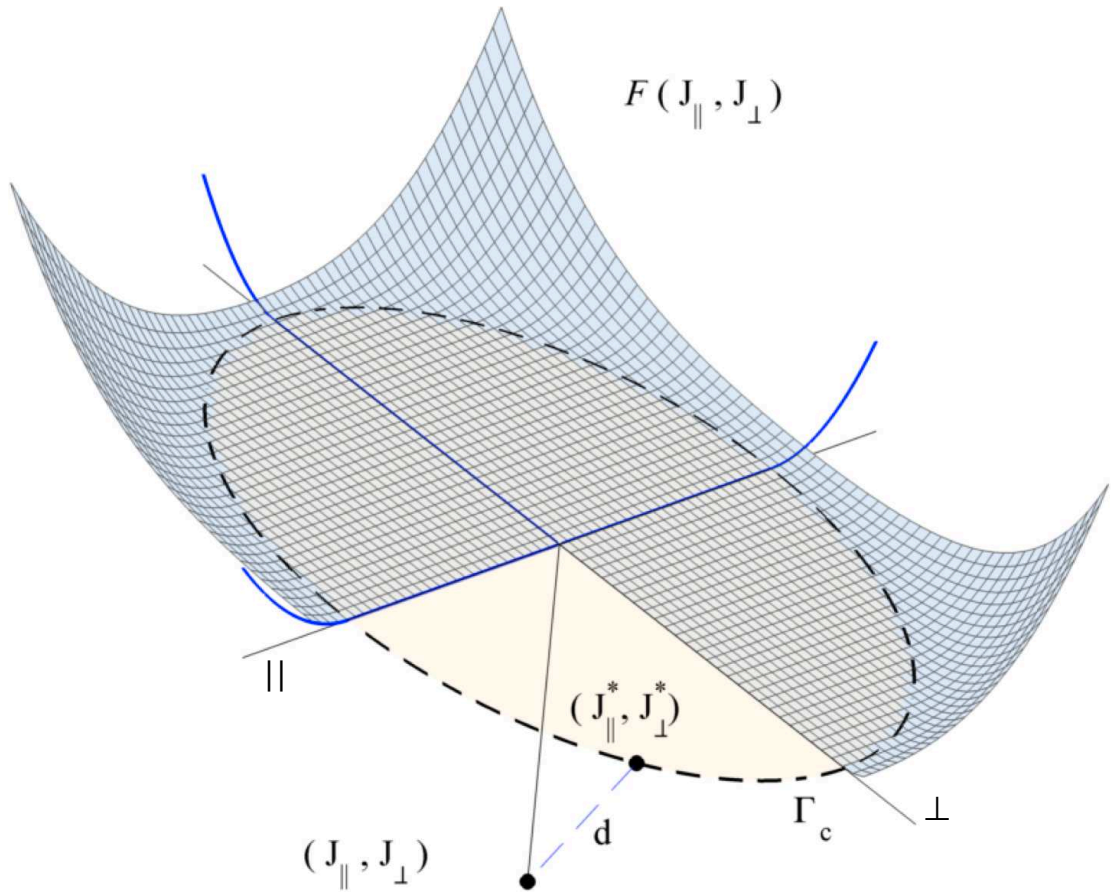}}
\caption{\label{Fig_2}(Color online) Dissipation function for a type-II superconductor close to the critical state. The horizontal plane is defined by the current density vector components either along ($J_{\parallel}$) or perpendicular ($J_{\perp}$) to the local magnetic field. For a given value of the current density vector $(J_{\parallel},J_{\perp})$, ${\cal F}$ is defined through the distance $d$ to the critical region boundary $\Gamma_{c}$. The related critical point ${\bf J_{c}\equiv (J_{\parallel}^{*},J_{\perp}^{*})}$ is also shown.}
\end{center}
\end{figure*}

Once more, a very general argument, uniqueness, will help. Let us consider the kind of elliptical region mentioned before, that is physically meaningful as related to dissipation mechanisms in either $J_{\parallel}$ or $J_{\perp}$ relative to the local magnetic field orientation. As depicted in Fig.\ref{Fig_2}, irreversible energy dissipation corresponding to some overcritical point ${\bf J}\not\in\Delta$ may be uniquely quantified by means of $d$, i.e.: the minimum distance from the point ${\bf J}$ to the boundary $\Gamma_c$. Geometrically, this means that expansion is done perpendicular to $\Gamma_c$. Analytically, one has to find some expression that allows to obtain the critical point ${\bf J_{c}}$ for each value of ${\bf J}$. In the case of the elliptic region considered here, this entails to solve a quartic equation. Of special mention is that meaningful cases\cite{general} as the isotropic model ($\Delta$ is a circle) and the Double Critical State Model ($\Delta$ is a rectangle) produce trivial conditions for the determination of the point ${\bf J_{c}}$.

%
%

From the physical point of view, the above mathematical conditions imply that for small perturbations around equilibrium, the induced electric fields will keep perpendicular to the levels of constant dissipation ${\cal F}$. This extends our previous result\cite{general} that just ``on surface" (critical state) ${\bf E}$ is normal to the boundary of the region $\Delta$. At least for small perturbations, a consistent theory that assumes the existence of a critical region with boundary $\Gamma_c$ may be obtained by using ${\bf E}\perp\Gamma_c$ and prolong this condition towards the resistive state. We call the readers' attention that this general property is not restricted to the case of a region $\Delta$ defined in terms of $J_{\parallel}$ and $J_{\perp}$. This would just fit the case of a homogeneous type-II superconductor, whose anisotropy only depends on the local magnetic field. Other physical scenarios, such as microstructure induced anisotropy are apparently tractable, just by using the required axes. For instance, one could use $J^{ab}$ and $J^{c}$ in the case of uniaxial symmetry for the current flow. $\Delta$ would be a cylinder with its axis parallel to the $\hat{\bf c}$ direction. Also, one could introduce a dependence of the critical current density on the orientation of the local magnetic filed relative to some axis in the material along which pinning is favored, i.e.: define $\Gamma_{c}$ by a function $J_{c}(\alpha)\equiv J_{c}(\hat{\bf H}\cdot\hat{\bf n})\equiv J_{c}(\cos{\alpha})$. In fact, the critical state solutions of such statements were already investigated in Ref.\cite{bifurcation} and the extension to the flux flow regime may be done in the above terms.

Before pursuing a specific mathematical statement of the above concepts, some clarifying words on the physical system are still due. On the one side, we have invoked the molecular thermal forces so as to describe the dissipative behavior beyond equilibrium. On the other side, criticality relates to a multiple number of possible ``equilibrium'' states for the system. The full framework is as follows. Indeed, thermal forces acting on the long term would eventually drive the system to a true (and unique) thermodynamic equilibrium state with ${\bf J}=0$. However, as we are focused on small  perturbations around the critical state, we just consider the fast relaxation towards a given metastable configuration determined by the applied fields and pinning forces. This corresponds to recalling that, in hard superconductors, much creep occurs in the first instances when the flux density gradient (basically, our $J_c$ parameter) is high \cite{tinkham}. Subsequent flux homogenization by flux hopping can be neglected in the magnetization processes under consideration as far as its typical time would be much beyond the experimental period.

\subsubsection{Thermodynamically admissible {\bf E}({\bf J}) laws: mathematical issues}~\label{Sec_223}

The dissipation function ${\cal F}$ must be defined, positive and obviously single valued, outside the 
region $\Delta$ of (sub)critical current densities, in the $\bf J$  vector space

\begin{equation}
{\cal F} : \IR^3/\Delta \to \IR ^+ \quad {\cal F}({\bf J}) \geq 0
\end{equation}
A Taylor expansion around the boundary of critical currents $\Gamma _c$ will encode the main
properties of the electromagnetic behavior close beyond the critical state, i.e., the  ${\bf E}({\bf J})$ constitutive law. 

\begin{figure*}
\begin{center}
{\includegraphics[width=0.4\textwidth]{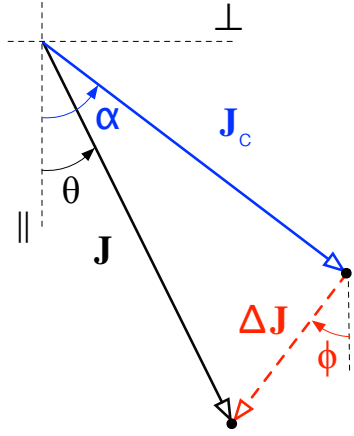}}
\caption{\label{Fig_3}(Color online) Schematics of the current density vectors in the over-critical state problem (as in Fig.\ref{Fig_2}). ${\bf J}$ stands for the actual current density,  ${\bf J}_c$ for the nearest critical current density vector, and $\Delta{\bf J}$ for their difference. Related angles are also defined.}
\end{center}
\end{figure*}

To start with, we will consider spatially isotropic samples. Thus, the surface $\Gamma _c$ will be axially 
symmetric around the local $\bf H$ direction, as well as symmetric under reflection through the 
plane normal to $\bf H$. $\Delta$ is also supposed convex. With this symmetry, the local 
electric field takes value in the plane defined by $\bf H$ and $\bf J$. Hereafter, we restrict the analysis to this plane, and use cartesian coordinates $(J_1,J_2)$, $(E_1,E_2)$ 
and $(H_1,0)$.
We note that the cartesian coordinates are nothing but the physical directions $(\parallel ,\perp)$ defined in Figs.\ref{Fig_2} and \ref{Fig_3}. However, in order to avoid awkward notation, within this section we use $(1,2)\equiv(\parallel ,\perp)$.

$\Gamma _c$ is then a closed 
curve surrounding the convex region of subcritical currents. In polar coordinates $(J_1,J_2)=(J \cos (\theta ),
J \sin (\theta ))$, and $\Gamma _c$ is determined by a function $J_c(\theta )$.

%
%

Let us see how the concept of perpendicular expansion introduced above arises. In principle, there is no particular point at $\Gamma _c$ from which to perform the Taylor expansion. 
Let us then consider an arbitrary ${\bf J}_c^{(1)}\in \Gamma _c$, and denote by 
$T({\bf J};{\bf J}_c^{(1)})$ the Taylor expansion of ${\cal F} \simeq T({\bf J};{\bf J}_c^{(1)})$  around this point $J_c^{(1)}$. 
It is reasonable to take a point ${\bf J}_c^{(1)}$ close to the value $\bf J$ of interest, in particular, nearest point  to $\bf J$ on the critical surface. However, a well behaved function ${\cal F}$ should admit 
compatible approximations from nearby points, say, ${\cal F} \simeq T({\bf J};{\bf J}_c^{(2)})$.
Therefore

\begin{equation}
T({\bf J}_0;{\bf J}_c^{(1)}) \simeq T({\bf J}_0;{\bf J}_c^{(2)}) 
\simeq T({\bf J}_0;{\bf J}_c(\theta )) \qquad \Rightarrow \qquad
\partial _{\theta} T= 0
\end{equation}
i.e., when fixing a value ${\bf J}_0$, the  derivative of $T({\bf J}_0;{\bf J}_c)$ along the curve $\Gamma_c$ must vanish. 

Now, for our purposes it suffices to include second and third order terms

\begin{eqnarray*}
T({\bf J};{\bf J}_c^{(1)}) &=& \frac {1}{2} [
\rho _{11}(J_1-J_{1c}^{(1)})^2 + \rho _{22}(J_2-J_{2c}^{(1)})^2 +  \\
 &+& 2 \rho _{12}  (J_1-J_{1c}^{(1)})(J_2-J_{2c}^{(1)}) ] + \\
&+& \frac {1}{6} [ \gamma _{111} (J_1-J_{1c}^{(1)})^3 +
3\gamma _{112} (J_1-J_{1c}^{(1)})^2(J_2-J_{2c}^{(1)}) +  \\
 &+& 3\gamma _{122} (J_1-J_{1c}^{(1)})(J_2-J_{2c}^{(1)})^2 +
\gamma _{222} (J_2-J_{2c}^{(1)})^3  ]
\end{eqnarray*}
with

\begin{equation}
\rho _{ij} = \frac {\partial ^2{\cal F}}{\partial J_i\partial J_j} ({\bf J}_c^{(1)})
\qquad 
\gamma _{ijk} = \frac {\partial ^3{\cal F}}{\partial J_i\partial J_j\partial J_k} ({\bf J}_c^{(1)})\, .
\end{equation} 
From the condition $\partial _{\theta} T= 0$ we get, to first order
 \begin{equation}
(\rho _{11}, \rho _{12}) \cdot {\bf T}(\theta ) = 0
\quad (\rho _{12}, \rho _{22}) \cdot {\bf T}(\theta ) = 0
\end{equation}
so that ${\bf T} = \partial _{\theta}(J_c(\theta ) \cos (\theta ),
J_c(\theta ) \sin (\theta ))$, the  tangent vector to $\Gamma _c$, belongs to the kernel of the resistivity tensor, which in physical terms means that no resistive losses occur for paths along  $\Gamma _c$

\begin{equation}
\Omega = \left( \begin{array}{cc}
\rho _{11} & \rho _{12} \\
\rho _{12} & \rho _{22}
\end{array} \right)
\end{equation}
 
On the other hand, to second order

\begin{eqnarray*}
(\gamma _{111}, \gamma _{112}) \cdot {\bf T}(\theta ) &=& - \partial _{\theta}\rho _{11}\\
(\gamma _{112}, \gamma _{122}) \cdot {\bf T}(\theta ) &=& - \partial _{\theta}\rho _{12}\\
(\gamma _{122}, \gamma _{222}) \cdot {\bf T}(\theta ) &=& - \partial _{\theta}\rho _{22}
\end{eqnarray*}
Note that for the strict second order Taylor polynomial ($\gamma_{ijk}=0$), the compatibility condition implies that $\Omega$ is  
constant, but this is no longer true when third order terms are considered. 

Now, if we use the vector basis $\{ \hat {\bf t},
\hat {\bf n} \}$, i.e.: unit tangent and normal vectors at a given point of $\Gamma _c$, the tensor $\Omega$ becomes

\begin{equation}
\Omega = \left( \begin{array}{cc}
0 & 0 \\
0 & \rho _{n}
\end{array} \right)
\end{equation}
Thus, the Hessian of ${\cal F}$ is positive semidefinite, with $\hat {\bf t}$ null eigenvector,
and $\rho _{n}$ the eigenvalue of $\hat {\bf n}$.

Remarkably, one can maintain the simplicity of the second order approach while allowing anisotropic 
$\Omega (\theta )$, by 
performing the Taylor expansion as follows; from each point ${\bf J}_c \in \Gamma _c$ we 
compute the Taylor expansion ``exclusively" along the normal halfline ${\bf J}_c + 
\Delta J \hat {\bf n}$. In such a way, and taking into account the convexity of the 
base curve, there is a univocal correspondence of each $\bf J$ with a particular 
Taylor expansion, and the whole region $\IR^2/\Delta _c$ is covered. The second order approach to the dissipation function becomes

\begin{equation}
{\cal F}({\bf J}) = \frac {1}{2} \rho (\alpha ) (\Delta J)^2
\quad {\bf J} = {\bf J}_c(\alpha ) + \Delta J \hat {\bf n} (\alpha )
\end{equation}
The correspondence ${\bf J}\equiv (J, \theta) \leftrightarrow (\Delta J, \alpha)$ is the geometric 
condition of nearest point of the curve $\Gamma _c$ to $\bf J$, and it obviously depends on the 
specific $J_c(\alpha)$ function. Note the distinction between $\theta$ and $\alpha$, the 
angular coordinates of $\bf J$ and ${\bf J}_c$ (Fig.\ref{Fig_3}).

Eventually, the electric field (${\bf E} = \nabla _{\bf J} {\cal F}$) is given by the polar coordinate expression

\begin{equation}
{\bf E} = \partial _J {\cal F}{\hat {\bf J}} + \frac {1}{J}\partial _{\theta} {\cal F}
{\hat {\bf \theta}} = E_{\parallel} {\hat {\bf J}}  + E_{\perp} {\hat {\bf \theta}}\, .
\end{equation}
For practical purposes, an intrinsic coordinate system representation will be useful. Thus, a better adapted expression in polar--like coordinates $\{ \Delta J, \phi \}$, with
$\Delta J$ the distance of $\bf J$ to $\Gamma _c$, and $\phi$ the angle between $\Delta{\bf J}$ and the ${\bf H}$ axis reads 

\begin{equation}
{\bf E} = \partial _{\Delta J} {\cal F}{\hat {\Delta \bf J}} + \frac {1}{l_\phi}
\partial _{\phi} {\cal F}
{\hat {\bf \phi}} \, .
\end{equation}
Here $l_{\phi}$ is the length of $\partial _{\phi}{\bf J}
= \partial _{\phi}{\bf J}_c + \Delta J \partial _{\phi}{\hat {\bf n}}$, and 
${\hat {\Delta \bf J}} \equiv {\hat {\bf n}}$. Then, one has

\begin{equation}
\label{Eq_E_intr}
{\bf E} = \rho (\phi ) (\Delta J){\hat {\bf n}} + \frac {1}{2 l_\phi}
\partial _{\phi} \rho (\Delta J)^2
{\hat {\bf \phi}}
\end{equation}
Notice that, in case of a constant $\rho$, the electric field is always parallel to ${\hat {\bf n}}$. However, 
as it will be analyzed below, in connection with voltage-current experiments, the ratio $E_{\parallel}$ and $E_{\perp}$ changes with
the separation of the working point and the critical region $\Delta {\bf J}={\bf J}-{\bf J}_{c}$ 
(see section \ref{Sec_32}).

%
\begin{figure*}[h]
\begin{center}
{\includegraphics[width=0.5\textwidth]{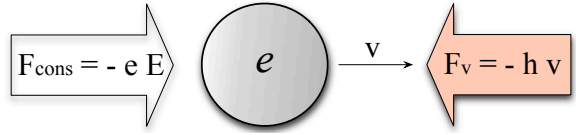}}
\caption{\label{Fig_4}Schematics of a simple irreversible system: a charge carrier subjected to the action of an electric field and a simultaneous viscous force.}
\end{center}
\end{figure*}
%

\subsubsection{Variational formulation of the conduction problem}~\label{Sec_224}

It has been argued that classical mechanics and its methods do not provide a complete framework for the analysis of electrical conduction. However, having exploited the consequences of the second law of thermodynamics, one can reconsider the problem. In particular, below we show that  by including the concept of admissible dissipation function ${\cal F}$ one can issue a variational formulation for the magnetic diffusion problem between steady conduction states. In this case, a unified description that encompasses normal conductors and type-II superconductors is presented.

Just with illuminating purpose we start by considering a 1D problem with a charged particle subject to an electric field ${E}$ with associated potential energy $U$. Recall that the Lagrangian formulation of Hamilton's principle is as follows

\vspace{1em}
\centerline{$\displaystyle{{\sf S}\equiv\int L(x,v)\, dt = \int \left[\frac{mv^2}{2}-U(x)\right]\, dt}$}
 
 \vspace{.5em}
\centerline{{Min {\sf S}}$\quad\Rightarrow\quad\displaystyle{\frac{d}{d t}\frac{\partial{L}}{\partial v}-\frac{\partial{L}}{\partial x}=0\quad\Leftrightarrow\quad}m\frac{d v}{dt}=-\frac{\partial U}{\partial x}\equiv F_{cons}$}

\vspace{1em}
\noindent where $F_{cons}$ stands for the {\em conservative} force $F_{cons}=q E$.

Consider now that a viscous drag acts on the particle (as depicted in Fig.\ref{Fig_4} for the case of an electron). Notice that a minimum principle leading to the sound equations of motion can still be formulated for a modified Lagrangian as shown below

\vspace{1em}

\centerline{$\displaystyle{\hat{\sf S}\equiv\int \hat{L}(x,v,t)\, dt  \quad{\rm with}\quad \hat{L}=L+{\cal F}t\equiv L+\frac{1}{2}hv^{2}\,t}$}

\vspace{.5em}

\centerline{${{\rm Min}\; \hat{\sf S}}\quad\Rightarrow\quad\displaystyle{\frac{d}{d t}\frac{\partial{\hat{L}}}{\partial v}-\frac{\partial{\hat{L}}}{\partial x}=0\quad\Leftrightarrow\quad m\frac{d v}{dt}\simeq-\frac{\partial U}{\partial x}-hv= F_{cons}+{F_{v}}}$}

\vspace{1em}
What has been done in deriving such formulation is to neglect variations of the viscous force $F_{v}$ within the interval of time considered. Then, the suggested approximation will be valid if minimization is applied ``iteratively'' with intervals of duration much less than the characteristic time $\tau\equiv h/m$. This relates to the so-called {\em Adiabatic Hypothesis} used in other physical disciplines:
\begin{itemize}
\item[ ] {\em If energy, though not conserved, varies slowly according to some parameter, then one is allowed to assume a kind of isolated system within small enough intervals of the temporal evolution.} 
\end{itemize}
An eventual calculation will help us to gain some more insight. Let us use the classical mechanics expression for calculating the energy of the particle in terms of the modified Lagrangian. Upon neglecting the term $\Delta F_v \Delta x$ the energy within a given interval is
\begin{equation}
\label{Eq_energy}
\displaystyle{{\cal E}=v\frac{\partial \hat{L}}{\partial v}-\hat{L}\simeq\frac{m v^2}{2}+\frac{1}{2}F_{v}{v}\Delta t}\, ,
\end{equation}
that is to say, we are effectively dealing with an isolated system which stores an average energy of one half of the full loss $F_{v}{v}\Delta t$.

Let us now see how the above arguments may be exported to the problem of electrical conduction. Within the quasisteady approximation, the incremental time-averaged field Lagrangian for conducting materials reads

\begin{equation}
\label{Eq_quasi_var}
\langle\hat{L}\rangle =\displaystyle{\frac{\mu_{0}}{2}\int_{\,\rthree}\;\|{\bf H}_{\rm n+1} -
{\bf H}_{\rm n} \|^{2} dV +\int_{\rm Vol}\;{\Delta t}\,{\cal F}dV}\equiv \langle\int_{\,\rthree}\;\hat{\cal L}\, dV\rangle
\end{equation}
with ${\cal F}$ the dissipation function introduced in Sec.\ref{Sec_22}. Here, ${\bf H}_{\rm n} $ means the magnetic field distribution at a given time instant given by $t_n$ and $\Delta t \equiv t_{n+1}- t_n$.

For the case of a normal conductor, it is relatively simple to show that the Euler-Lagrange equations lead to the desired diffusion equation (Eq.(\ref{Eq_diff})). Thus, if one assumes a diagonal resistivity matrix, i.e.: $\Omega^{ij}=\rho_0\delta^{ij}$ and replaces ${\cal F}$ by its value, it follows

\begin{eqnarray}
\quad\quad\frac{\partial\hat{\cal L}}{\partial {\bf H}_{\rm n+1}}&=&\sum_{j}\frac{\partial}{\partial x_j}\frac{\partial\hat{\cal L}}{\partial(\partial {\bf H}_{\rm n+1}/\partial x_j)}
\nonumber\\
&\Downarrow&
\nonumber\\
\mu_0\frac{{\bf H}_{\rm n+1}-{\bf H}_{\rm n}}{\Delta t}&=&-\rho_0\nabla\times\nabla\times{\bf H}_{\rm n+1}=\rho_0\nabla^2{\bf H}_{\rm n+1} 
\end{eqnarray}
i.e.: the time discretized form of Eq.(\ref{Eq_diff}).

Outstandingly, the interest of Eq.(\ref{Eq_quasi_var}) is that it gives way to the possibility of applying direct numerical minimization methods. In particular, this allows to deal with non-simple forms of ${\cal F}$ as required for the investigation of type-II superconductors (recall Fig.\ref{Fig_2}).

From the mathematical point of view, one can distinguish two kinds of problems as related to the minimization of the functional in Eq.(\ref{Eq_quasi_var}): (a) strictly variational structures, when either the first or the second term may be neglected, and (b) a quasi-variational structure when both are relevant \cite{Berdichevski}. Physically, one would speak about:

\begin{itemize}

\item[(i)] Equilibrium-like processes, leading to the critical state (the dissipation function may be neglected and one basically obtains the configuration that balances magnetic and pinning forces).
\item[(ii)] Steady states within a dissipative regime (the magnetic inertia $\Delta {\bf H}$ may be neglected, typically because external sources are fixed).
\item[(iii)] Quasisteady evolutions when both inertia and dissipation have to be included (system {\em diffuses} towards the critical state  or towards a new steady state if dictated by the external sources).

\end{itemize}

The rest of this paper will be focused on several examples in which the properties of type-II superconductors close to  the critical state are calculated based on the theory issued in this section. The presentation will be organized according to the three cases described above.


\section{Numerical applications}~\label{Sec_3}
As it was explained in Sec.\ref{Sec_2}, the core of our theoretical proposal is the existence of a dissipation function ${\cal F}$ that one may write down as a quadratic expression of the macroscopic current density vector components. Recall that both from the mathematical point of view, and also as concerns the physical background, ${\cal F}$ should be defined as a certain distance $d$. Actually, $d$ is a measure of the separation between the operation point (given by  the value of the current density ${\bf J}$) and a certain equilibrium value ${\bf J}_{c}$ that depends on ${\bf J}$ itself. The specific form of the function ${\cal F}$ depends on the critical current region boundary $\Gamma_c$, as we have explicitly shown in Fig.\ref{Fig_2} for the case of an elliptic behavior. 

In this section, we present a number of examples in which the above ideas have been applied to various practical configurations. Let us start by writing down some useful equations. Just for clarity, our statements refer to the elliptical region in Fig.\ref{Fig_2}, but generalization is apparent.

\begin{itemize}

\item Notice that one may classify a given value of ${\bf J}$ (in components $(J_{\parallel},J_{\perp})$) as lying inside $\Gamma_c$ or not, by calculating the sign of the quantity
\begin{equation}
\delta_{\Gamma}({\bf J}) ={\left(\frac{J_{\parallel}}{J_{c\parallel}}\right)^{2}+\left(\frac{\vphantom{J_{\parallel}}J_{\perp}}{\vphantom{J_{c\parallel}}J_{c\perp}}\right)^{2}-1} \, ,
\end{equation}
i.e.: $\delta_{\Gamma}({\bf J})$ is either negative or positive when ${\bf J}$ is either within or beyond the contour of $\Gamma_c$ (and null for the {\em critical} values ${\bf J}_{c}\in\Gamma_c$).

\item The distance function for a given point $(J_{\parallel},J_{\perp})$, also involving its image ${\bf J}_{c}\equiv (J_{\parallel}^{*},J_{\perp}^{*})$ is given by
\begin{equation}
d^{\,2} =  \left(J_{\parallel}-J_{\parallel}^{*}\right)^{2}+ \left(\vphantom{J_{\parallel}}J_{\perp}-J_{\perp}^{*}\right)^{2} \, .
\end{equation}
Here, $J_{\parallel}^{*}$ and $J_{\perp}^{*}$ have to be determined for the actual critical current law (region) under consideration. In the case of an elliptic model, the criterion of minimum distance between $(J_{\parallel},J_{\perp})$ and the boundary leads to solving a quartic equation.

\item For a given value $(J_{\parallel},J_{\perp})$, the dissipation function ${\cal F}({\bf J})$ may be written as 

\begin{equation}
{\cal F}= \frac{1}{2}\,\rho\, \Theta_{\Gamma}({\bf J})\, d^{\,2}({\bf J})
\end{equation}
with $\rho$ the material resistivity, and $d$ the distance function referred above. $\Theta_{\Gamma}$ stands for a step function whose value is zero within $\Gamma_c$ and one over the outside. A useful representation is
\begin{equation}
\Theta_{\Gamma}({\bf J})=\frac{1+\tanh{[k \delta_{\Gamma}({\bf J})]}}{2} \; , \; k\gg 1 \, .
\end{equation}
\end{itemize}
Recall that for the rather usual experimental configuration in which ${\bf J}={\bf J}_{\perp}$ (the components of {\bf J} parallel to the magnetic field are zero), the above formulation may be cast as follows (mind the thick lines in Fig.\ref{Fig_2}):
\begin{eqnarray}
\delta_{\Gamma}&=&(J_{\perp}/J_{c\perp})^{2} -1
\nonumber\\\nonumber\\
J_{\perp}^{*}&=&\left\{
\begin{array}{l l}
J_{c\perp} & \qquad \qquad {\rm if}\; J_{\perp}> J_{c\perp}\\
-J_{c\perp} & \qquad \qquad {\rm if}\; J_{\perp}< -J_{c\perp}
\\
\end{array}\right.
\nonumber\\\nonumber\\
d^{\,2}&=&\left\{
\begin{array}{l l}
(J_{\perp}-J_{c\perp})^{2} & \quad\;\; {\rm if}\; J_{\perp}> J_{c\perp}\\
(J_{\perp}+J_{c\perp})^{2} & \quad\;\; {\rm if}\; J_{\perp}< -J_{c\perp}
\\
\end{array}\right.
\end{eqnarray}

%
\begin{figure*}[!]
\begin{center}
{\includegraphics[width=0.6\textwidth]{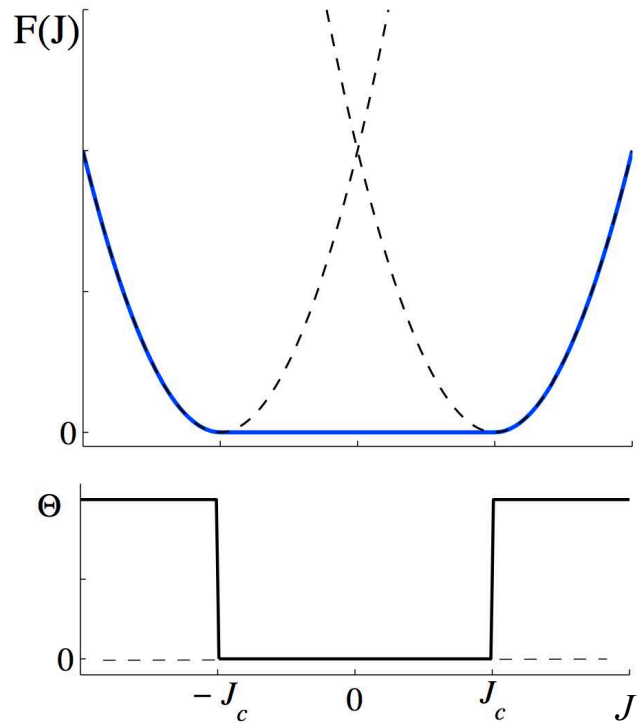}}
\caption{\label{Fig_5}(Color online) Detail of a one-dimensional dissipation function for type-II superconductors. ${\cal F}$ is built upon multiplying two parabolae by the step functions depicted below and adding them. Note that the horizontal axis, i.e.: $J$ is common to boh plots.}
\end{center}
\end{figure*}
%

These relations mean that, for a vast number of experimental setups, ${\cal F}$ may be built by composing two parabolae and a step function as illustrated in Fig.\ref{Fig_5}. The examples supplied below are based on this assumption. For the readers' sake, we eventually provide a useful discrete form of Eq.(\ref{Eq_quasi_var}) that is based upon the transformation of the electromagnetic problem in terms of potentials\cite{badia_ajp} and on identifying the set of elementary circuits related to the problem's symmetry, viz.

\begin{eqnarray}
\label{Eq_diff_discr}
\hspace{-1cm}{\displaystyle\qquad\quad\qquad\frac{\mu_{0}}{2}\int_{\,\rthree}\;\|{\bf H}_{\rm n+1} -
{\bf H}_{\rm n} \|^{2} dV}& \qquad\quad\int_{\rm Vol}\;{\Delta t}\,{\cal F}dV
\\
\hspace{-2.5cm}{\langle\hat{L}\rangle_{\rm dis} =} \overbrace{\frac{1}{2} \sum\limits_{i,j} I_i {M}_{ij} I_j - \sum\limits_{i,j} {\tilde I}_i {M}_{ij} I_j
+ \sum\limits_i I_i \left( A^e - {\tilde A}^e\right)}&+ \overbrace{\frac{1}{2} \Delta t \sum\limits_i R _i \Theta (\pm I_{c_i})\left(I_i \mp I_{c_i}\right)^{2}}
\nonumber
\end{eqnarray}

\noindent Here $ A^e $ means the vector potential component along the current lines, 
related to the applied magnetic field, $ {M}_{ij}$ denotes the inductance between elementary currents $I_i$, $I_j$ flowing 
along the specific circuits of the problem, and  the tilded quantities concern previous time layer. $\Delta t$ is the incremental time of our calculation, $R_i$ stands for the incremental resistance of the $i$-th circuit and $I_{c_i}$ its critical current. Application to different problems will imply different expressions for the matrix elements ${M}_{ij}$, the applied vector potential, the resistance and the critical currents. 

Just for completeness, we mention that other possibilities for the dissipation function may be explored as it will be shown below. In fact, the customary power-law expression for the $E(J)$ law will be also investigated in this paper, through the related dissipation function

\begin{equation}
\displaystyle{\int_{\rm Vol}\; {\Delta t}\,{\cal F}dV\;\to  F_{0}\Delta t\sum\limits_i \left(\frac{I_i}{I_{c_i}}\right)^{m}}
\end{equation}

\subsection{Metastable equilibrium: {the critical states}}~\label{Sec_31}

To start with, we present an application of minimizing the expression in Eq.(\ref{Eq_diff_discr}) in a genuine ``critical" situation, i.e.: the dissipation term is not included explicitly, but replaced by the constraint $J \leq J_{c}$ in the minimization of the first term. Physically, this corresponds to the conventional critical state problems in which overcritical excursions are neglected. We have chosen a case of interest, that has been solved analytically, thus allowing a straightforward comparison to our numerical results. In connection with the widely used configuration of a flat sample in a perpendicular magnetic field, we apply our method to the circular disk geometry. According to Mikheenko and Kuzovlev\cite{mikheenko}, the sheet current density distribution in a superconducting disk subject to a uniform perpendicular field of intensity $H_0$ is given by

\begin{equation}\label{Eq_disk}
J^{s}(r)=\left\{
\begin{array}{ll}
{\displaystyle\frac{2J^{s}_{c}}{\pi} \tan ^{-1}{\left[\frac{(r/R)\sqrt{R^{2}-a^{2}}}{\sqrt{a^{2}-r^{2}}}\right]}}& \qquad  r\leq a
\\
J^{s}_c &  \qquad a \leq r\leq R \, ,
\end{array}
\right.
\end{equation}
with $a\equiv R/\cosh{(2 H_{0}/J_{c}d)}$. $d$ stands for the (small) thickness of the disk and $R$ for its radius. Here, $J^{s}$ denotes the so-called sheet current that averages the current density over the thickness, i.e.: $J^{s}(r)\equiv \int\, J(r,z) dz$.

\begin{figure*}
\begin{center}
{\includegraphics[width=0.5\textwidth]{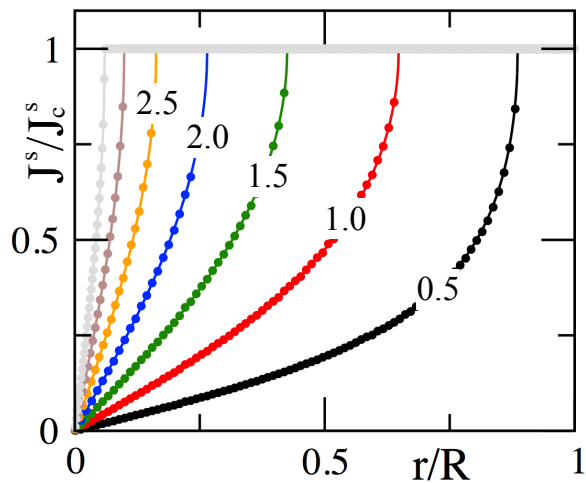}}
\caption{\label{Fig_6}(Color online) (Sheet) current density profiles in the equilibrium (critical) states that are induced in a superconducting disk by an increasing magnetic field along its axis (given by $2H_{0}/J_{c}d = 0.5, 1, 1.5, 2, 2.5, 3, 3.5$ from the outermost to the innermost curve). Lines correspond to the analytical solution in Ref.\cite{mikheenko} (see text) and symbols to our numerical results.}
\end{center}
\end{figure*}

Fig.\ref{Fig_6} displays the comparison of our results and those obtained from the above equation. We stress the fact that numerics produce the correct solution, even for situations in which the variable does not reach the true critical value. Recall that, physically, the strict inequality ($J^{s}(r)<J^{s}_c$) relates to the meaning of the sheet current as an average over the sample thickness.  This situation occurs when a non-penetrated core still exists somewhere within the sample, i.e.: at some points $(r_{0},z)$ one has $J(r_{0},z)=J_c$ for the peripheral values of $z$ and $J(r_{0},z)=0$ in the central core.

%
%

\subsection{Steady states: {the voltage criterion}}~\label{Sec_32}

Below, we analyze the opposite limit of the general conduction problem with type-II superconductors. The steady situation in which a dissipative state is maintained by the external action (current source) will be analyzed. Thus, based on the considerations introduced in Sec.\ref{Sec_223} for the law ${\bf E}=\nabla_{{\bf J}}{\cal F}$, we concentrate on recently reported voltage-current experiments \cite{clem}, specifically designed to investigate the complex ${\bf E}({\bf J})$ law and underlying critical boundary $\Gamma_c$ in High-T$_c$ superconductors. Owing to the smart design of the experiment, a direct analytical study is allowed.

First, we recall that the results in Ref.\cite{clem} show that the elliptic model for $\Gamma_c$ gives an excellent
fit to measurements. The authors have also obtained the polar angular dependence of the angle between ${\bf E}$  and ${\bf J}$ , say $\beta(\theta)$ (see Fig.\ref{Fig_3} for definitions and, in our case, consider Eq.(\ref{Eq_E_intr}) in order to obtain the direction of ${\bf E}$ relative to $\Delta{\bf J}$). 
In their experimental configuration $\hat {\bf J}$ is fixed along the film direction, while $\hat {\bf B}$ is 
applied at different angles $\theta$. The criterion for reaching and exceeding the critical state comes from a 
fixed value of ${\bf E}$ parallel to the direction of ${\bf J}$ ($E_{\parallel}\equiv{\bf E}\cdot \hat{\bf J} = E_0$). 

From the elliptical curve of critical currents

\begin{equation}
{\displaystyle \left( \frac {J_{\parallel}}{J_{c\parallel}}\right)^2 + \left( \frac {\vphantom{J_{\parallel}}J_{\perp}}{\vphantom{J_{c\parallel}}J_{c\perp}}\right)^2   = 1}
\end{equation}
we get the parametric representation

\begin{equation}
J_c(\alpha )^2 \left[\left( \frac {\cos (\alpha)}{J_{c\parallel}}\right)^2 + \left( \frac 
{\sin (\alpha)}{J_{c\perp}}\right)^2 \right]  = 1
\end{equation}
in terms of which the normal vector is given by $(\cos (\alpha)/J_{c\parallel}^2,
\sin (\alpha)/J_{c\perp}^2)$. Recalling the notation introduced in Sec.\ref{Sec_223}, $\hat {\bf n} = (\cos (\phi), \sin (\phi))$, we have 
$\gamma ^2\tan (\phi ) = \tan (\alpha )$, with the anisotropy ratio $\gamma = J_{c\perp}/J_{c\parallel}$.

Now, the ratio $E_{\perp}/E_{\parallel}$ for small dissipation ($\Delta J \to 0$) is basically that of the 
electric field components at $\Gamma _c$, where ${\bf E} \perp \Gamma _c$ ($\beta = \phi - \theta$), .i.e.:

\begin{eqnarray}
\label{Eq_Phic}
E_{\perp}/E_{\parallel} & = \tan (\theta -\phi ) = 
\frac {\tan (\theta )-\tan  (\phi)}{1 + \tan (\phi)\tan (\theta )}
= \frac {\tan (\theta)(\gamma ^2 -1)}{\gamma ^2 + \tan (\theta )^2} 
\nonumber\\
&\equiv \Phi _c(\gamma ,\theta)
\end{eqnarray}
because in this limit $\alpha = \theta$.

In order to quantify the influence of the dissipation parameters in the electric field ratio, we proceed by considering a first order correction in terms of $\Delta J$. In the simplest case of constant $\rho$ (see Sec.\ref{Sec_223}) this maintains 
${\bf E} \parallel {\hat {\bf n}}$, but now $\alpha \neq \theta$. From the trigonometric relations in the triangle of 
sides $\bf J$, ${\bf J}_c$ and $\bf {\Delta J}$
and the condition $E_{\parallel} = E_0$ we get $\theta - \alpha \simeq \Phi _c E_0/(\rho J_c)$, and then

\begin{eqnarray}
\label{Eq_PhiDj}
E_{\perp}/E_{\parallel} &\approx   
\Phi _c \left[ 1 - \delta \gamma ^2 (1 + \Phi _c ^2)
\frac {(1 + \tan ^2(\theta))^{1/2}(\gamma ^2 + \tan ^2(\theta ))^{1/2}}
{\gamma ^4 + \tan ^2(\theta )} \right]
\nonumber\\
 &\equiv \Phi _{\Delta J}(\gamma ,\theta,\delta)
\end{eqnarray}
where $\delta = E_0/(\rho J_{c\perp})$ is the small parameter in the expansion.

Eventually, we consider the most general case in which anisotropic resistivity is allowed. Let us assume the form $\rho = \rho _{\parallel } \cos ^2(\theta ) + \rho _{\perp }\sin ^2(\theta)$, and $r^2 \equiv\rho _{\parallel}/\rho _{\perp}$ the anisotropy ratio.
Starting with Eq.(\ref{Eq_E_intr}) and after some algebra we get

\begin{eqnarray}
\label{Eq_PhiR}
E_{\perp}/E_{\parallel}&\approx  
 \Phi _{\Delta J} + \delta \Phi _r (\Phi _{\Delta J}^2 - 1) \sqrt{1 + \Phi _{c}^2}
\frac{(\gamma ^2 + \tan ^2(\theta ))^{3/2}}
{(r^2 + \tan ^2(\theta ))(\gamma ^4 + \tan ^2(\theta ))^{1/2}}
\nonumber\\
&\equiv \Phi _{cr}(\gamma ,\theta,\delta,r)
\end{eqnarray}
with $\Phi _r \equiv(1-r^2)\tan (\theta )/(r^2 + \tan ^2(\theta))$.

%
\begin{figure*}
\begin{center}
{\includegraphics[width=.9\textwidth]{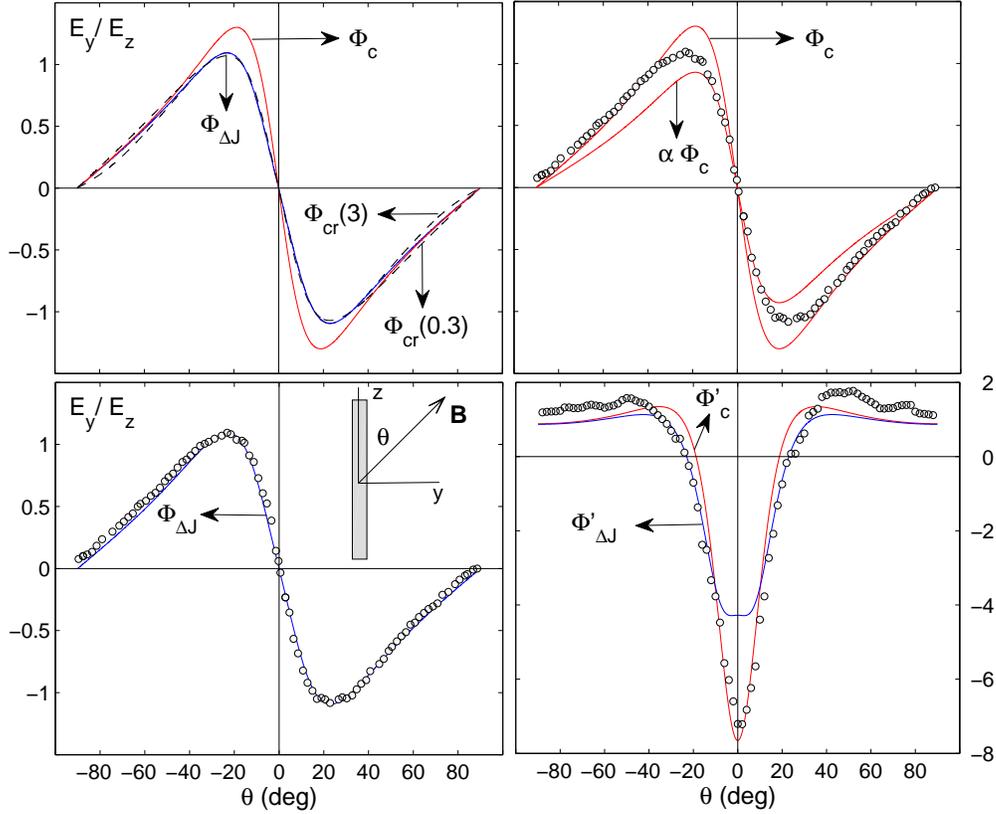}}
\caption{\label{Fig_7}(Color online) Plots of the electric field ratio $E_{y}/E_{z}$ in the magnetotransprt experiment sketched in the inset. A transport current is directed along the $z$ axis of a superconducting wire, that is also subjected to an applied magnetic field (induction at an angle $\theta$). Electric fields are calculated for $J\gtrsim J_c$ according to Eqs.(\ref{Eq_Phic},\ref{Eq_PhiDj},\ref{Eq_PhiR}) and plotted as lines. Symbols come from the experiment in Ref.\cite{clem}: YBCO film at $T=85 K$. Top left: comparison between our theoretical curves $\Phi_{c}$, $\Phi_{\Delta J}$, and $\Phi_{cr}$ for a couple of values of the anisotropy parameter ($r = 0.3\; \&\; r=3$). Top right: $\Phi_{c}$ is compared to the experimental data, either multiplied or not by a factor $\alpha$. Bottom left: $\Phi_{\Delta J}$ is compared to the experimental data. Bottom right: comparison of the derivatives of the function $[E_{y}/E_{z}](\theta)$ obtained from theory and experiment. See the text for details about parameter values in our simulations and scaling of data.}
\end{center}
\end{figure*}
%

Notice that, considering that the anisotropy ratio $\gamma$ is given, our theory includes two parameters, $ \delta$ and $r^2$, that may be used to fit the experiment. Ultimately this will either allow to obtain the resistivity coefficients or (if they are known by other means) check for consistency.

Our theory has been used to analyze the experimental information in Ref.\cite{clem}. Fig.\ref{Fig_7} shows our predictions as compared to the available data. Basically, we plot the angular dependence of the ratio $E_{\perp}/E_{\parallel}$ as given by the different levels of approximation in Eqs.(\ref{Eq_Phic},\ref{Eq_PhiDj},\ref{Eq_PhiR}), i.e.: $\Phi_{c}$, $\Phi_{\Delta J}$ and $\Phi_{cr}$. Our theoretical curves are obtained by using $\gamma = 0.34$ the value that is reported by the authors as a direct information from their experiments. Several facts are of mention:

\begin{itemize}

\item Experimental data have been ``recalibrated", multiplying by a factor of 0.714

\item If one takes for granted that the value of $\gamma $ is correct, $\Phi_c$ does not produce the best fit even when multiplying by constant factors (in the plot, we have the direct data and also multiplied by $\alpha = 0.714$). On the contrary, allowing a constant factor between theory and experiment, the best results are given by  $\Phi_{\Delta J}$ with $\delta = 0.15$.

\item the correction introduced by anisotropic resistivity ($\Phi _{cr}$) does not seem relevant to this experiment. Lacking direct information, we just see that changing its value from $3$ to $1/3$ only minor effects are observed.

\item If one analyzes the first derivative of $\Phi$ and compares to the results obtained from the experimental data, some doubts arise, because $\Phi_{\Delta J}^{'}$ does not seem to show the real shape, whereas $\Phi_{c}^{'}$ looks more realistic.

\end{itemize}

From these issues, we state that the comparison between theory and experiment is reasonable. First, from the value $\delta = 0.15$ one obtains a resistivity: $\rho\approx 10^{-9}\,\Omega\, m$, which is nor far from the typical values for flux flow resistivities at the experimental range (around $10^{-8}\,\Omega\, m$ for YBCO films at $T=85\, K$). Second, the calibration factor for the experimental data ($C=0,714$) is not unreasonable because the electric fields were just obtained from the averaged expression $voltage/distance$. According to the authors of Ref.\cite{clem} this procedure should be ``refined'' when (as it is the case in parallel configurations) the electric field structure along the sample is not homogeneous, a fact that is under investigation.

\subsection{Transient behavior: {relaxation towards the critical state}}~\label{Sec_33}

\begin{figure}[!]
\begin{center}
{\includegraphics[width=0.7\textwidth]{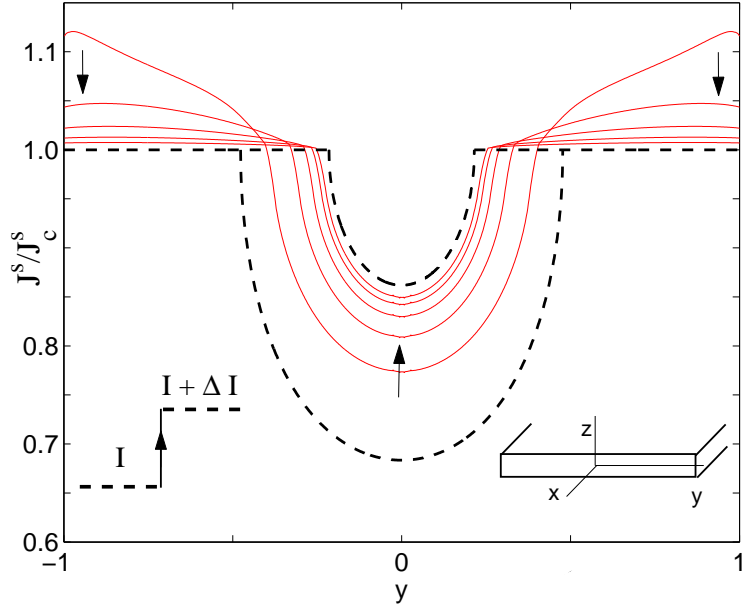}}
\caption{\label{Fig_8}(Color online) Illustration of the diffusion processes simulated in this work. Here, a step is applied to the transport current $I$ along the $X$-axis of a thin strip (as shown in the insets). The circulating sheet current $J(y)$ increases over the initial critical state profile (lower dashes) and then ``relaxes" towards the subsequent critical state (upper dashes). Continuous lines correspond to the intermediate current density profiles, and arrows are used to indicate the time evolution of $J$ at different parts of the sample. These relaxation curves are obtained subsequent to the step in the transport current ($I\to I+\Delta I$) and occur at equidistant time intervals of value $0.4\,\tau_{0}$.}
\end{center}
\end{figure}

Finally, based on the minimization of the expression in the Eq.(\ref{Eq_diff_discr}), we put forward several cases that explicitly show the diffusion of electromagnetic fields in a hard superconductor. We will concentrate on the tape geometry (Fig.\ref{Fig_8}) and analyze relaxation towards the critical state, when either transport current or magnetic field steps are applied. By considering different values of the bias characteristic period $\tau_{0}$ as compared to the magnetic diffusion time constant $\tau_{\rho}=\mu_{0}L^{2}/\rho$ (with $L$ the sample's dimension), we will display different conditions in which critical states are (or aren't) realized after relaxation. It will be shown that the conditions studied may be applied to establish validity of the CSM approximation in AC experiments.

The general features of our simulations are shown in Fig.\ref{Fig_8}. Notice that we obtain the current density profiles that occur between two equilibrium states when a step of transport current is introduced. Mainly, we have studied the influence of the ratio $\tau_{0}/\tau_{\rho}$ on the diffusion process within the quadratic dissipation function framework developed in the previous sections. This relates to the piecewise linear behavior of the $E(J)$ law. Furthermore, for completeness, we have also analyzed the results for power-law relations as indicated above.

%
%
\newpage

%
\begin{figure*}[h]
\begin{center}
{\includegraphics[width=1.0\textwidth]{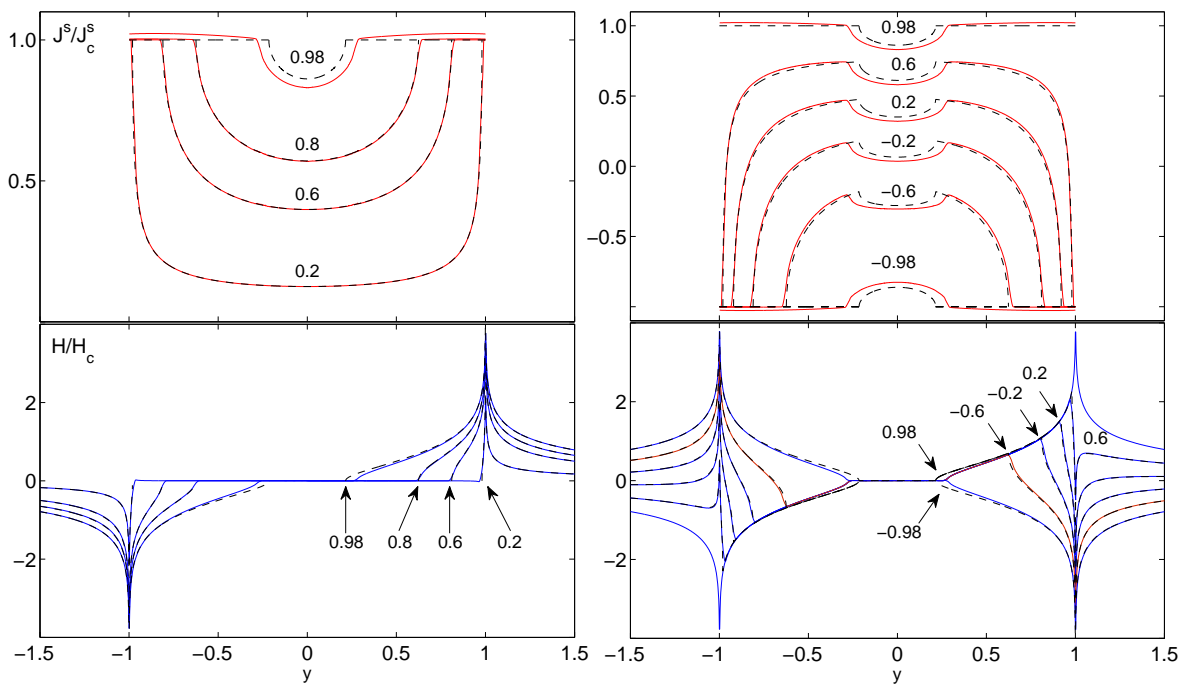}}
\caption{\label{Fig_9}(Color online) Normalized sheet current (top) and magnetic field (bottom) profiles in a superconducting strip carrying an impressed  transport current $I$, that is cycled as labelled in the curves (see text for the definition of units). Dashed lines correspond to the analytical solutions in Ref.\cite{brandt}, and continuous lines to our numerical diffusion calculations. In this case, and for each current interval, we just plot the relaxation profiles obtained for a quadratic dissipation function at the time $t=\tau_{0}=2.0\,\tau_{\rho} $, previous to the subsequent step in transport current. The panels to the left stand for the increasing branch of the applied current, while the right panels display the decreasing branch profiles.}
\end{center}
\end{figure*}
%

\subsubsection{Superconducting strips with transport current}~\label{Sec_331}

\begin{figure*}
\begin{center}
{\includegraphics[width=1.05\textwidth]{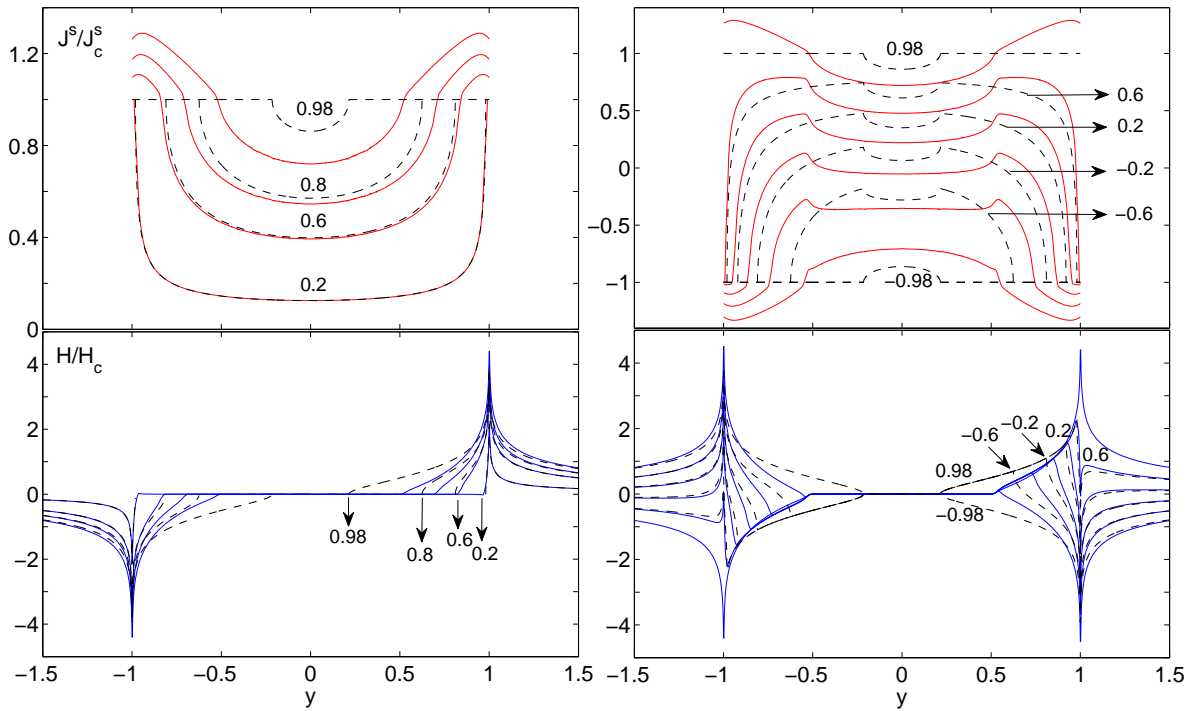}}
\caption{\label{Fig_10}(Color online) Same as Fig.\ref{Fig_9}, but now for $\tau_{0}/\tau_{\rho} =0.2$.}
\end{center}
\end{figure*}

A number of situations related to the experimental setup sketched in Fig.\ref{Fig_8} have been studied, but only a representative set of results are shown below (Figs.\ref{Fig_9} and \ref{Fig_10}). To start with, the choice of the long strip (tape) geometry allows a straightforward comparison to analytical results for the related limiting critical states. Thus, the thick dashed lines in that figure (critical states) have been obtained from the well-known expressions for the above defined sheet current\cite{brandt}

\begin{equation}
\label{Eq_J_BI}
J^{s}(y)=\left\{
\begin{array}{ll}
{\displaystyle\frac{2J^{s}_{c}}{\pi} \tan ^{-1}\sqrt{\frac{a^{2}-b^{2}}{b^{2}-y^{2}}}}& \qquad  |y|\leq b
\\
J^{s}_c &  \qquad b \leq |y|\leq a \, ,
\end{array}
\right.
\end{equation}
with $a$ the tape ``half-thickness" and $b=a\surd{(1-I^{2}/I_{\rm max}^{2}})$, $I_{\rm max}\equiv 2aJ^{s}_{c}\equiv 2\pi a H_{c}$. In our case, normalized units have been used for all the physical quantities: $J/J_{c}, H/H_{c}, I/I_{\rm max}, y/a$. The  magnetic field profiles have been obtained by integration of the current density. Analytical expressions derived from Eq.(\ref{Eq_J_BI}) may be found in Ref.\cite{brandt}. As for the results of this paper, straightforward numerical integration of the current density data was performed. 

Fig.\ref{Fig_9} shows the penetration profiles for a transport current that increases from the virgin state until the value $I/I_{\rm max}=0.98$ is reached. Then, a negative ramp towards the value $I/I_{\rm max}=-0.98$ was applied. The associated field penetration profiles are shown in the lower panels. Just for the readers' sake, we recall that expressions for critical state curves in the negative ramp of transport current may be obtained from Eq.(\ref{Eq_J_BI}) by applying linear superposition\cite{brandt}. For the numerical curves in this plot, the calculated relaxation takes place under the condition $\tau_{0}/\tau_{\rho}=2$ and here we just show the profiles previous to the application of the following step (evolution takes place in the fashion shown in Fig.\ref{Fig_8}). Notice that for the increasing branch, a high degree of coincidence between the ``relaxed'' profiles and the exact critical states displays. It is only for the nearly penetrated sample that relaxation is noticeably incomplete at $t=\tau_{0}=2.0\,\tau_{\rho} $. A mismatch appears in the central part of the tape, that is inherited by the decreasing branch of the cycle, i.e.: when  new steps are applied before reaching the equilibrium profile. As a counterpart, a reasonable coincidence is always observed at the sample edges. Apparently, this behavior relates to the concept that flux penetrates from the surface, and that dissipation is an integral over the sample, thus giving place to a most effective relaxation when a smaller region is involved. In other words, relaxation leads to equilibrium quicker close to the surface than in the bulk.

%
%
Fig.\ref{Fig_10} has been obtained in a similar fashion, but under the condition $\tau_{0}/\tau_{\rho}=0.2$. As expected, clearly incomplete relaxation is obtained. In practice, this would lead to magnetization profiles with absolute values beyond the critical state limit, and also to higher AC losses, corresponding to the noticeable excursions of the transport current density towards the region $J>J_{c}$. 

We conclude this part by mentioning that results obtained for the power law relation are rather similar to the previously described behavior, when the respective choices $m=100$ and $m=10$ are done (to be compared to the cases $\tau_{0}/\tau_{\rho}=2$ and $\tau_{0}/\tau_{\rho}=0.2$ respectively).

%
\begin{figure*}[b]
\begin{center}
{\includegraphics[width=1.05\textwidth]{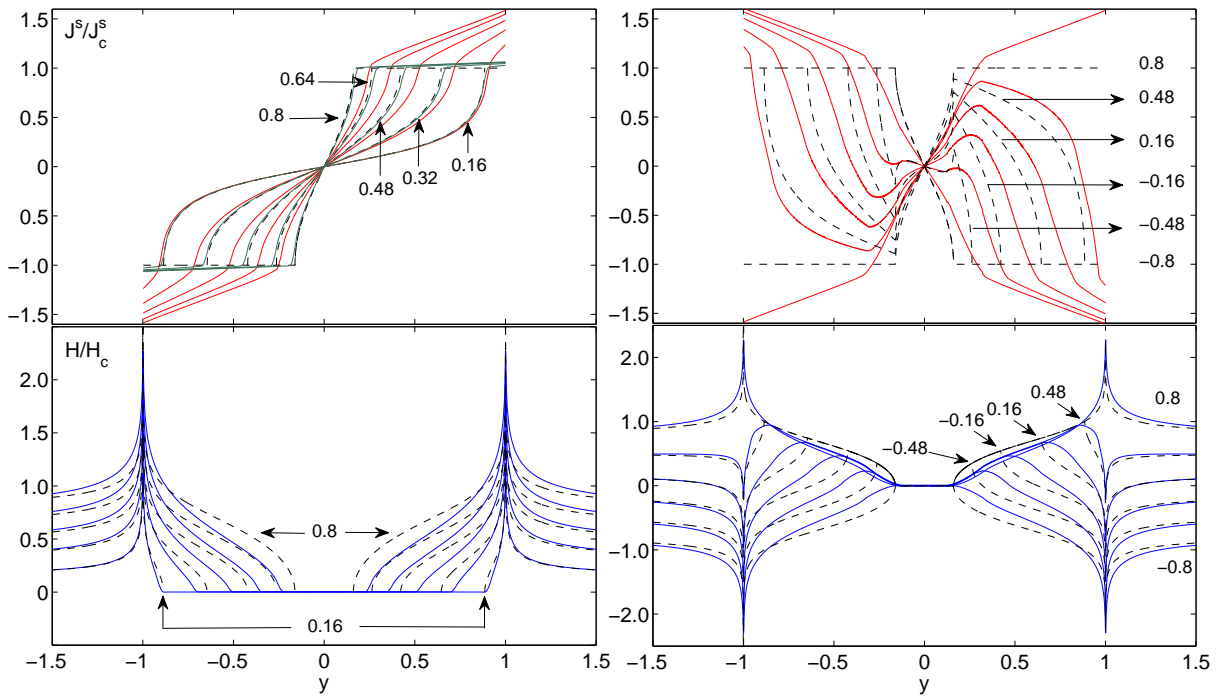}}
\caption{\label{Fig_11}(Color online) Normalized sheet current (top) and magnetic field (bottom) profiles for a superconducting strip in an impressed perpendicular magnetic field $H_{0}$. To the left, the depicted profiles correspond to the values  $H_{0}/H_{c}= 0.16, 0.32, 0.48, 0.64, 0.8$. To the right, the applied field is cycled, i.e.: $H_{0}/H_{c}= 0.8, 0.48, 0.16, -0.16, -0.48, -0.8$. Dashed lines correspond to the analytical solutions in Ref.\cite{brandt}, and continuous lines to our numerical diffusion calculations. In this case, we plot the profiles obtained for a quadratic dissipation function with $\tau_{0}/\tau_{\rho} =0.2$ just previous to the subsequent step in transport current. Notice that the upper left plot also includes the results for $\tau_{0}/\tau_{\rho} =2.0$, that are omitted in the rest to avoid confusion, and virtually coincide with the analytical lines. Labeling is bypassed in some apparent cases.}
\end{center}
\end{figure*}
%
\subsubsection{Superconducting strips with applied magnetic field}~\label{Sec_332}

\begin{figure*}[h]
\begin{center}
{\includegraphics[width=1.05\textwidth]{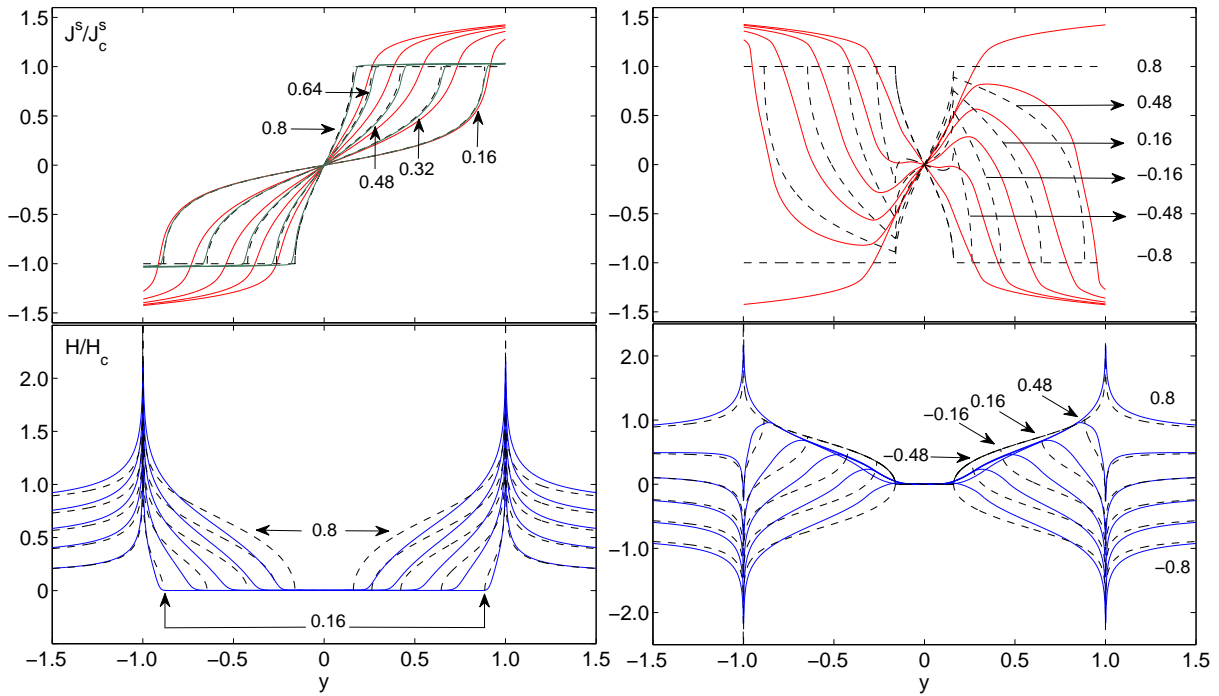}}
\caption{\label{Fig_12}(Color online) Same as Fig.\ref{Fig_11} but for the power law model with $m=10$ and $m=100$.}
\end{center}
\end{figure*}

The diffusing current and field profiles in a thin strip in a perpendicular field $H_0$ and with zero transport current have been calculated under a wide set of conditions. Figs.\ref{Fig_11} and \ref{Fig_12} present the main features observed in our simulations. In this case, the seed for the analytical evaluations is the expression

\begin{equation}
\label{Eq_J_BI_H}
J^{s}(y)=\left\{
\begin{array}{ll}
{\displaystyle\frac{2J^{s}_{c}}{\pi} \tan ^{-1}\frac{c y}{\sqrt{{b^{2}-y^{2}}}}}& \qquad  |y|\leq b
\\
{\displaystyle J^{s}_c \frac{y}{|y|}}&  \qquad b \leq |y|\leq a \, ,
\end{array}
\right.
\end{equation}
where $c\equiv \tanh (H_{0}/H_{c})$. Magnetic field profiles may be obtained by integration and negative ramp equations by linear superposition\cite{brandt}.
Also as before, dashed lines stand for the analytical results of equilibrium profiles, and continuous curves for our numerical calculations of finite time (incomplete) relaxation processes. We only display the profiles corresponding to the diffusion step just previous to the change of external condition.

Fig.\ref{Fig_11} shows the results for the quadratic dissipation function, and Fig.\ref{Fig_12} for the power-law relation. It is noticeable that, in both cases, a remarkable degree of coincidence between the relaxed profiles and the eventual critical state solution occurs for $\tau_{0}/\tau_{\rho}\geq 2$. Nevertheless, important differences appear for the case  $\tau_{0}/\tau_{\rho}\simeq 0.2$. As related to such differences, an important feature has been observed. We call the readers' attention that in Fig.\ref{Fig_11} $J^{s}(y)$ displays a basically linear penetration profile close to the sample's edge, whereas a kind of square root trend is observed in Fig.\ref{Fig_12}. This may be explained as follows. According to Faraday's law, one has $\partial_{y} E_{x}=\partial_{t}B_{z}$. Then, for small variations $\partial_{t}B_{z}$ will be practically constant in space and so that $E_{x}\sim(\partial_{t}H_{0})\,y$, which implies a linear behavior of $J^{s}_{x}(y)$ in the case of a piece-wise linear $E(J)$ law, and an inverse power law when $E\sim J^n$ and then, $J^{s}_{x} \sim y^{-n}$.
%
%
%
%
\begin{figure*}[h]
\begin{center}
{\includegraphics[width=.6\textwidth]{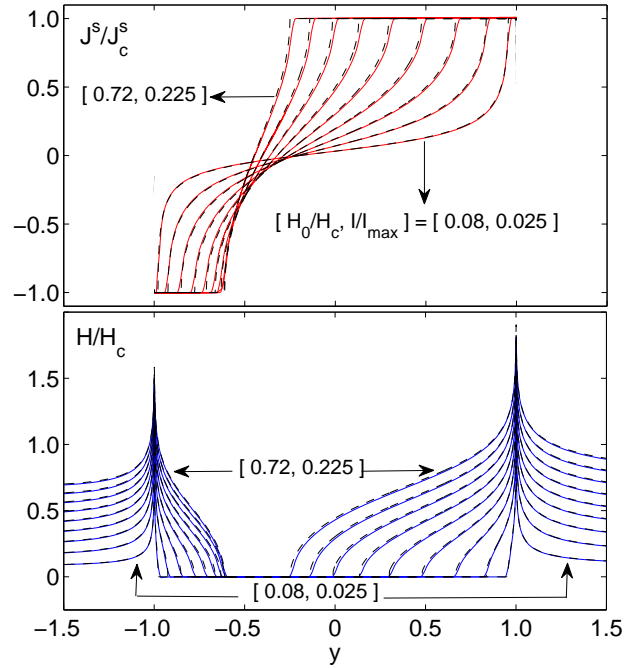}}
\caption{\label{Fig_13}(Color online) Normalized sheet current (top) and magnetic field (bottom) profiles for a ``field-like" state in superconducting strip with both transport current and perpendicular magnetic field applied with constant ratio. In normalized units, the different steps correspond to $H_{0}/H_{c}=0.08, 0.16, ... 0.72$ and $I/I_{\rm max}=0.025, 0.05, ... 0.225$. Dashed lines correspond to the analytical solutions in Ref.\cite{brandt}, and continuous lines to our numerical diffusion calculations. In this case, we plot the profiles obtained for a quadratic dissipation function with $\tau_{0}/\tau_{\rho} =20$.}
\end{center}
\end{figure*}

\subsubsection{Superconducting strips with transport current and applied field}~\label{Sec_333}

Just for completeness, we will also provide an example of relaxation towards the critical state when both a transport current and a magnetic field are applied to the superconducting strip. Analytical expressions for the case of synchronous ramps of current and field have also been provided in Ref.\cite{brandt}. Thus, the solution is built through the ``generating function''

\begin{equation}
\label{Eq_J_BI_H_I}
j(y,a,b)=\left\{
\begin{array}{ll}
1 &  \qquad b \leq y\leq a 
\nonumber\\
{\displaystyle\frac{1}{\pi} \cot ^{-1}\frac{b^{2}-ay}{p}}
& \qquad  |y|\leq b
\\
0&  \qquad -\infty \leq y\leq -b\, ,
\end{array}
\right.
\end{equation}
where ${\displaystyle p\equiv\sqrt{(y^2-b^2)(a^2-b^2)}}$. Explicitly, one has

\begin{equation}
J^{s}(y)=J^{s}_{c}\,[\, j(y+w,a+w,b)+p\, j(-y-w,a-w,b)]
\end{equation}
with the definition $w\equiv (I/I_{\rm max})/\tanh (H_{0}/H_{c})$.

Fig.\ref{Fig_13} shows the comparison of our numerical results and the analytical calculation described above. Following the nomenclature in Ref.\cite{brandt} we have concentrated on a so-called {\em field-like} situation given by $r\equiv I/2\pi a H_{0}$= 0.32. 

In this case, we have considered a high relaxation ratio $\tau_{0}/\tau_{\rho} =20$, and a practical coincidence with the critical state profiles is observed.



\section{Conclusions}~\label{Sec_4}

The motivation of this work has been to explore the possibility of extending the critical state concept so as to include the effects of flux flow resistance in a general sense, i.e.: allowing the possibility of overcritical current density in several dimensions. More specifically, we have studied the case of homogeneous type-II samples, in which anisotropy is introduced by the direction of the local magnetic field. However, generalization to other cases, such as microstructure induced anisotropy are straightforwardly dealt with. Conceptually, we have chosen a macroscopic (thermodynamic) point of view, that avoids the explicit consideration of the underlying vortex physics. In brief, our theory relies on two facts: (i) the experimental evidence that in type-II superconductors dissipativeless currents are allowed when the components of ${\bf J}$ either parallel or perpendicular to some specific direction do not exceed specific thresholds, and (ii) the resistive transition has to occur according to the laws of entropy production. These considerations are formulated in a geometrical language through the definition of the so-called {\em dissipation function}, that takes values over the space of current densities, i.e.: ${\cal F}({\bf J})$.  ${\cal F}$ goes to zero within the so-called critical region ${\Delta}$ (generalization of the one dimensional condition $J\leq J_c$) and is a positive definite quadratic form beyond. The main role of this function relates to the issue of the material law ${\bf E}({\bf J})$ that appears by just imposing consistency and uniqueness in the physical quantities. In fact, one has ${\bf E}=\nabla_{\bf J} {\cal F}$.

The relevant approximations of the theory may be stated in terms of time constants. As related to the typical time scale of the macroscopic experiments (i.e.: $1 {\rm Hz}< \omega < 1 {\rm kHz}$), the following scale applies: (i) charge recombination processes are assumed to occur instantly (magnetoquasistatic approximation), (ii) magnetic diffusion occurs very fast in the Critical State limit and may be modulated by a time constant $\tau_{\rho}\propto 1/\rho$ in our theory, and finally (iii) thermal activation and relaxation to the true equilibrium (${\bf J}=0$) occurs very slowly and may be neglected.

From the practical point of view, and taking advantage of the variational interpretation of the electromagnetic problem\cite{badia_2001}, we put forward a minimization statement that gives way to the numerical form of our theory. Eq.(\ref{Eq_diff_discr}) displays the function that is minimized, and is the central result of this article. One can identify two basic contributions that relate to the physics of the problem: (i) the inertial terms that account for the reversible energy storage, and (ii) the energy dissipation term that includes ${\cal F}$. Inspired by this interpretation, we have worked out a number of examples that illustrate the application of the theory either straightforwardly to the standard critical state problem, to steady states in which permanent dissipation is forced by some external action, or to the transient (diffusive) processes that occur in between successive critical states, or between steady states as dictated by modified external sources.

Based on the comparison of our results to other theoretical works on the critical state, we conclude that the complex non-linear diffusion processes that take place according to the theory, converge to the previously reported critical states solutions for a given set of external conditions. Excellent agreement is observed when the system is allowed to relax within the typical excitation period. 

Related to the analysis of dissipative steady states, we stress the fact that our theory unifies the standard CSM framework and the results of current-voltage techniques, used to derive the critical current parameters. In this sense, we give a number of relations (Eqs.(\ref{Eq_Phic}) to (\ref{Eq_PhiR})) that allow to analyze the multicomponent ${\bf E}({\bf J})$ measurements designed for characterizing the critical current behavior of High-T$_{c}$ superconductors\cite{clem}. Along this line, we have compared our predictions to those experiments and can conclude that, although a reasonable agreement is obtained, more experimental information would be desirable in order to choose between one theory or another for the overcritical states.

Although we state that a quadratic dissipation function (and thus a piecewise linear ${\bf E}({\bf J})$ law) is the more judicious choice for investigating  the behavior close to the critical state, the rather extended use of a power-law relation has also been checked against our results with good degree of coincidence. Only, second order qualitative differences can be observed. Notice, in passing, that big discrepancies with the critical state are out of reach of our theory, but also of experiments because holding current densities much above the critical value would result in a blow-up of the sample.

Further work along the lines of this paper entails the application of the theory to obtain the relaxation profiles in higher dimensional systems, such as structurally anisotropic materials and problems with multicomponent magnetic fields with non homogeneous electric field profiles.


\section*{Acknowledgements}

Funding of this research by Spanish MINECO and the European FEDER Program (Projects MAT2011-22719, ENE2011-29741) and by Gobierno de Arag\' on (Research group T12) is gratefully acknowledged.

\vspace{0.5 cm}

%
%


\begin{thebibliography}{99}

\bibitem{bean} {Bean C P 1962  {\em Phys. Rev. Lett.} {\bf 8} 250; 1964  {\em Rev. Mod. Phys.} {\bf 36} 31}

\bibitem{pla} {Richardson R A, Pla O and Nori F 1994 {\em Phys. Rev. Lett.} {\bf 72} 1268}

\bibitem{surdacki} {Surdacki P 2003  {\em Phys. C} {\bf 387} 234}

\bibitem{cha03} {Cha Y S 2003  {\em IEEE Trans. Appl. Supercond.} {\bf 13} 2028}

\bibitem{clem} {Clem J R, Weigand M, Durrell J H and Campbell A M 2011  {\em Supercond. Sci. Technol.} {\bf 24} 062002}; {Campbell A M 2011  {\em Supercond. Sci. Technol.} {\bf 24} 091001}
\bibitem{orlando} {Orlando T P and Delin K A, 1991 {\em Foundations of Applied Superconductivity}, Prentice Hall, New Jersey}

\bibitem{landau} {Landau L D and Lifshitz E M 1984 {\em Electrodynamics of Continuous Media, Vol. 8 (\rm Course of Theoretical Physics)} Pergamon Press, Oxford}

\bibitem{goldstein} {Goldstein H, Poole C and Safko J 2001 {\em Classical Mechanics}, Addison Wesley Longman, San Francisco}

\bibitem{explanation} {Note: in Classical Mechanics, frictional forces of the form ${\bf F_v}=-h {\bf v}$ are derived from a function ${\cal F}$ (Rayleigh's dissipation function) defined by ${\cal F}\equiv(1/2)hv^{2}$ that relates to the system's energy loss by $d\epsilon /dt = -2{\cal F}$}

\bibitem{prigogine} {Prigogine I 1967 {\em Introduction to Thermodynamics of Irreversible Processes} North-Holland, Amsterdam}

\bibitem{klein} {Klein M J and Meijer P H E 1954  {\em Phys. Rev.} {\bf 96} 250}

\bibitem{general} Bad\'{\i}a  A,  L\'opez C and Ruiz H S 2009  {\em Phys. Rev. B} {\bf 80} 144509

\bibitem{bifurcation} Bad\'{\i}a  A and  L\'opez C  2002  {\em J. Appl. Phys.} {\bf 92} 6110

\bibitem{tinkham}   {Tinkham M 1980  {\em Introduction to Superconductivity} Robert E Krieger Publishing Company, Malabar, Florida} 

\bibitem{Berdichevski} {Berdichevski V L 1967 {\em Variational Principles of Quantum Mechanics} Interaction of Mechanics and Mathematics, Springer-Verlag, Heidelberg}

\bibitem{badia_ajp} {Bad\'{\i}a A 2006 {\em Am. J. Phys.} {\bf 74} 1136}

\bibitem{mikheenko} {Mikheenko P N and Kuzovlev  Y E 1993  {\em Phys. C} {\bf 204} 229}

\bibitem{brandt} Brandt E H and Indenbom M 1993  {\em Phys. Rev. B} {\bf 48} 12893

\bibitem{badia_2001} Bad\'{\i}a A and L\'opez C 2001 {\em Phys. Rev. Lett.} {\bf 87} 127004



\end{thebibliography}
\end{document}